\documentclass[twocolumn,showpacs,nofootinbib,preprintnumbers,amsmath,amssymb,showkeys,superscriptaddress]{revtex4}
\usepackage{epsfig}
\usepackage{bbold}
\usepackage{dsfont}
\usepackage{color}
\usepackage{graphicx}
\usepackage{slashed}
\usepackage{dcolumn}
\usepackage{bm}
\usepackage{mathtools}
\DeclarePairedDelimiter\bra{\langle}{\rvert}
\DeclarePairedDelimiter\ket{\lvert}{\rangle}
\DeclarePairedDelimiterX\braket[2]{\langle}{\rangle}{#1 \delimsize\vert #2}
\usepackage[final]{feynmp}
\makeatletter
\def\endfmffile{%
	\fmfcmd{\p@rcent\space the end.^^J%
		end.^^J%
		endinput;}%
	\if@fmfio
	\immediate\closeout\@outfmf
	\fi
	\ifnum\pdfshellescape=\@ne
	\immediate\write18{mpost \thefmffile}%
	\fi}
\makeatother

\newcommand{\QCD}{{\rm QCD}}
\newcommand{\FP}{{\rm FP}}
\newcommand{\tr}{{\rm tr\, }}

\newcommand{\cb}{\bar c}
\newcommand{\Ab}{\bar A}
\newcommand{\Db}{\bar D}
\newcommand{\psib}{\bar \psi}

\def\slashii#1{\setbox0=\hbox{$#1$}             
	\dimen0=\wd0                                 
	\setbox1=\hbox{\sl/} \dimen1=\wd1            
	\ifdim\dimen0>\dimen1                        
	\rlap{\hbox to \dimen0{\hfil\sl/\hfil}}   
	#1                                        
	\else                                        
	\rlap{\hbox to \dimen1{\hfil$#1$\hfil}}   
	\hbox{\sl/}                               
	\fi}                                         %
\def\slashiii#1{\setbox0=\hbox{$#1$}#1\hskip-\wd0\hbox to\wd0{\hss\sl/\/\hss}}
\newcommand{\beq}{\begin{equation}}
	\newcommand{\eeq}{\end{equation}}
\newcommand{\bea}{\begin{eqnarray}}
	\newcommand{\eea}{\end{eqnarray}}
\newcommand{\nn}{\nonumber \\}

\newcommand\eqn[1]{(\ref{#1})}      
\newcommand\Eqn[1]{Eq.~(\ref{#1})}  
\newcommand\Fig[1]{Fig.~\ref{#1}}  

\begin{document}
	
	
	\title{A perturbative study of the QCD phase diagram for heavy quarks\\ at nonzero chemical potential: two-loop corrections}

	\author{J. Maelger}
	\affiliation{Centre de Physique Th\'eorique, Ecole Polytechnique, CNRS, Universit\'e Paris-Saclay, F-91128 Palaiseau, France.\vspace{.1cm}}
	\affiliation{Astro-Particule et Cosmologie (APC), CNRS UMR 7164, Universit\'e Paris Diderot,\\ 10, rue Alice Domon et L\'eonie Duquet, 75205 Paris Cedex 13, France.\vspace{.1cm}}
	
	\author{U. Reinosa}%
	\affiliation{%
		Centre de Physique Th\'eorique, Ecole Polytechnique, CNRS, Universit\'e Paris-Saclay, F-91128 Palaiseau, France.\vspace{.1cm}}%
	
	\author{J. Serreau}
	\affiliation{Astro-Particule et Cosmologie (APC), CNRS UMR 7164, Universit\'e Paris Diderot,\\ 10, rue Alice Domon et L\'eonie Duquet, 75205 Paris Cedex 13, France.\vspace{.1cm}}

	\date{\today}

	\begin{abstract}

We extend a previous investigation \cite{reinosa2015perturbative} of the QCD phase diagram with heavy quarks in the context of background field methods by including the two-loop corrections to the background field effective potential. The nonperturbative dynamics in the pure-gauge sector is modelled by a phenomenological gluon mass term in the Landau-DeWitt gauge-fixed action, which results in an improved perturbative expansion. We investigate the phase diagram at nonzero temperature and (real or imaginary) chemical potential. Two-loop corrections  yield an improved agreement with lattice data as compared to the leading-order results. We also compare with the results of nonperturbative approaches. We further study the equation of state as well as the thermodynamic stability of the system at two-loop order. Finally, using simple thermodynamic arguments, we show that the behavior of the Polyakov loops as functions of the chemical potential complies with their interpretation in terms of quark and anti-quark free energies.

 \end{abstract}

\maketitle


\section{Introduction}
\label{sec:intro}
The phase diagram of QCD is the subject of intense theoretical and experimental investigations \cite{Stephanov:2007fk,Endrodi:2011gv,Stephanov:1999zu,Aggarwal:2010cw}. It is expected to present a rich structure, governed by unique fundamental features of the theory, such as confinement and chiral symmetry breaking \cite{Svetitsky:1985ye,Glashow:1967rx,Banks:1979yr}. On the one hand, this offers an original angle to study those features and, on the other hand, the phase structure of the theory may have nontrivial consequences in various situations of phenomenological interest, such as the early Universe or dense astrophysical objects. A large panel of theoretical methods have then been developed to explore the phase diagram of QCD matter in thermodynamic equilibrium at finite nonzero temperature $T$, baryon chemical potential $\mu$, magnetic field $B$, etc., ranging from lattice Monte Carlo simulations to various continuum approaches \cite{deForcrand:2002hgr,DElia:2002tig,DElia:2010abb,Borsanyi:2013bia,Braun:2007bx,Braun:2009gm,Fischer:2013eca,Fischer:2014vxa,Canfora:2015yia,Quandt:2016ykm,Reinhardt:2017pyr}. The former, when available, are capable of tackling the exact nonperturbative dynamics of the theory and are essentially limited by statistical errors. The latter, instead, directly access averaged quantities, such as correlation functions, but necessarily rely on some approximation scheme which has to be justified {\it a posteriori}. 

A major open question concerns the structure of the phase diagram in the $(T,\mu$)-plane and, in particular, the possibility of a first order transition line at low $T$, governed by the restoration of the chiral symmetry and ending at a critical point \cite{Fodor:2004nz}. This is a situation where standard Monte Carlo techniques fail because of a severe sign problem at nonzero real $\mu$ in QCD. Continuum approaches based on the QCD Lagrangian, instead, face the issue of properly including the relevant degrees of freedom, which, in the regime of interest, are not only quarks and gluons but also bound states \cite{Eichmann:2016yit,Braun:2014ata}. 

Another interesting line of investigation is to explore the phase structure of the theory also in parameter space, for instance, by varying the quark masses $M_f$, the number of colors $N$, etc. At large quark masses, chiral symmetry is strongly broken and the main phenomenon at work is confinement. The phase structure of the $N_c=3$ theory in this regime is summarized in the celebrated Columbia plot: in the infinite mass limit, the pure Yang-Mills (YM) theory presents a first order transition; for large but finite quark masses the transition weakens and eventually turns into a continuous transition for critical values of the masses and a crossover for lower masses. The location of the corresponding critical surface in the space of quark masses depends on the (possibly imaginary) chemical potential $\mu$ and is known to shrink towards larger masses as $\mu^2$ increases from negative to positive values. Furthermore, the theory presents a rich phase structure in the case of an imaginary chemical potential \cite{Roberge:1986mm}. Despite its seemingly academic nature, this problem is interesting as it can bring useful information on the real chemical potential case. This idea can be tested in the heavy quark limit since, in that case, the sign problem of Monte Carlo techniques at real $\mu$ can be avoided by a large mass expansion \cite{Fromm:2011qi}. Reproducing the rich phase structure of the theory in the heavy mass regime is a nontrivial task for continuum methods and a useful benchmark so as to whether typical approximation schemes correctly include the relevant dynamics. 

A series of recent works have investigated a new perturbative approach based on a simple modification of the Faddeev-Popov (FP) Lagrangian in the Landau and Landau-DeWitt (LDW) gauges, which amounts to adding a bare mass term for the gluon field \cite{Curci76,Tissier:2010ts,Tissier:2011ey}. This is motivated by two observations. First, lattice calculations in the Landau gauge show that the gluon propagator in the vacuum \cite{Bonnet:2000kw,Bonnet:2001uh,Cucchieri_08b,Bogolubsky09,Bornyakov09,Iritani:2009mp,Maas:2011se,Oliveira:2012eh} and at finite temperature \cite{Cucchieri11,Cucchieri12,Aouane:2011fv,Maas:2011ez,Silva:2013maa}  saturates at vanishing momentum, which amounts to a nonzero screening mass\footnote{The gluon propagator also presents a violation of spectral positivity, which shows that the corresponding massive-like excitation does not correspond to an asymptotic state, as expected from confinement. In the massive Curci-Ferrari model mentioned here, such a spectral positivity violation is generated by loop effects \cite{Tissier:2010ts}.}  in $d=3$ and $d=4$ dimensions. Including such a mass term in the Lagrangian on phenomenological grounds leads to a well-defined perturbative expansion and actual one-loop calculations of ghost, gluon, and quark two- and three-point functions yield good agreement with existing lattice data \cite{Tissier:2010ts,Tissier:2011ey,Pelaez:2013cpa,Pelaez:2014mxa,Pelaez:2015tba,Reinosa:2013twa}. This suggests that, at least in the Landau gauge, the residual interactions beyond those responsible for the generation of this effective mass term can be treated perturbatively. The second motivation stems from the fact that the FP gauge-fixing procedure ignores the Gribov ambiguities, which are, however, known to play a nontrivial role in the regime of infrared momenta \cite{Gribov77,Neuberger:1986vv,Neuberger:1986xz,Zwanziger89,Vandersickel:2012tz,Dudal08}. The massive extension of the FP Lagrangian is the simplest deformation of the gauge-fixed Lagrangian which preserves the ultraviolet (UV) properties of the theory \cite{Curci76,Tissier:2008nw}. 

The perturbative approach in the massive extension of the LDW gauge has proven extremely efficient in describing known aspects of the QCD phase diagram with heavy quarks at nonzero temperature and chemical potential. A simple one-loop calculation correctly captures the order of the confinement-deconfinement transition at nonzero $T$ in pure YM theories \cite{Reinosa:2014ooa}. In that case, two-loop corrections have also been computed \cite{Reinosa:2014zta,reinosa2016two}, which quantitatively improve the one-loop results for, say the transition temperature, and cure some (but not all) unphysical features of the one-loop results for thermodynamical observables. Heavy dynamical quarks have been discussed in Ref.~\cite{reinosa2015perturbative} where, again, a simple one-loop calculation reproduces the rich phase diagram mentioned above and produces numbers for the critical line in the Columbia plot in quantitative agreement with lattice data. 

The purpose of the present work is to complete this series of works by computing the two-loop correction from the quark sector. The main aim is to study whether, as was the case for the SU($2$) and SU($3$) pure YM theories, the two-loop corrections quantitatively improve on the one-loop results. This can be tested on a wide variety of results at vanishing, imaginary, and real chemical potential. We shall also use the two-loop results to study some thermodynamical aspects of the system.

The paper is organized as follows. In Sec.~\ref{sec:model}, we recall the general setting, that is, the massive extension of the LDW gauge. The Feynman rules in the appropriate canonical bases  and the detailed calculation of the two-loop quark-gluon sunset diagram, including renormalization, are detailed in Sec.~\ref{sec:calculation}. We present our results for the phase diagram and the thermodynamics in Sec.~\ref{sec:results} and we conclude in Sec.~\ref{sec:conc}. Technical details are gathered in the Appendices~\ref{Reduction}--\ref{sec:last}.

\section{Generalities}
\label{sec:model}

The Euclidean action of QCD in $d$ dimensions with $N$ colors and $N_f$ quark flavors reads 
\begin{equation}
\label{eq:action}
  S_\QCD=\int_x\bigg\{\frac 14 F_{\mu\nu}^aF_{\mu\nu}^a+\sum_{f=1}^{N_f}\psib_f( {\slashiii  {\cal D}} +M_f+\mu\gamma_0)\psi_f\bigg\}\,,  
\end{equation}
where we have defined $\smash{\int_x\equiv\int_0^\beta d\tau\int d^{d-1}x}$, with $\beta$ the inverse temperature, $\smash{F_{\mu\nu}^a\equiv\partial_\mu A_\nu^a-\partial_\nu A_\mu^a+g f^{abc}A_\mu^b A_\nu^c}$, with $g$ the bare coupling constant and $f^{abc}$ the structure constants of the SU($N$) group, and $\smash{{\cal D}_\mu\psi\equiv\left(\partial_\mu-igA^a_\mu t^a\right)\psi}$, with $t^a$ the generators of the group in the fundamental representation, normalized as $\smash{\tr t^at^b=\delta^{ab}/2}$. Finally, $\mu$ denotes the chemical potential. 

We leave the Dirac and color indices of the quark fields implicit and $\psi_f$ and $\psib_f$ are understood in the common sense as column and line bispinors, respectively. The Euclidean Dirac matrices $\gamma_\mu$ are hermitian\footnote{They are related to the standard Minkowski matrices as $\gamma_0\equiv\gamma_M^0$ and $\gamma_i\equiv-i\gamma_M^i$.} and satisfy the anticommutation relations $\{\gamma_\mu,\gamma_\nu\}=2\delta_{\mu\nu}$.

\subsection{The (massive) Landau-DeWitt gauge}
In the pure Yang-Mills case, the deconfinement transition is tantamount to the spontaneous breaking of center symmetry \cite{Svetitsky:1985ye}. To study the latter in a gauge-fixed setting, it is thus convenient to work in a gauge where the symmetry is manifest at each step. This for instance the case for the LDW gauge \cite{Braun:2007bx}, that we now recall.

The gauge field $A_\mu^a$ is decomposed into a background field $\Ab_\mu^a$ and a fluctuating contribution as
\begin{equation}
  \label{eq:Aba}
  A_\mu^a=\Ab_\mu^a+a_\mu^a\,,
\end{equation}
and the LDW gauge-fixing condition reads
\beq
  (\Db_\mu a_\mu)^a=0\,,
 \eeq
 with $\smash{\Db^{ab}_\mu\equiv\delta^{ab}\partial_\mu+ gf^{acb}\Ab_\mu^c}$ the background covariant derivative in the adjoint representation. The corresponding Faddeev-Popov gauge-fixing action reads
\begin{equation}
  \label{eq:fp}
  \begin{split}
    S_\FP=\int_x \Big\{(\Db_\mu \cb)^a(D_\mu c)^a+ih^a(\Db_\mu a_\mu)^a\Big\}\,,
  \end{split}
\end{equation}
where $\smash{D^{ab}\equiv\delta^{ab}\partial_\mu+ gf^{acb}A_\mu^c}$, $c$ and $\cb$ are anticommuting ghost fields and $h$ is a Lagrange multiplier, also known as the Nakanishi-Lautrup field. 

In practice, at each temperature, the background field $\Ab_\mu^a$ is chosen such that the expectation value $\langle a_\mu^a\rangle$ vanishes in the limit of vanishing sources. In the pure Yang-Mills case, such backgrounds are obtained as the absolute minima\footnote{See Ref.~\cite{reinosa2016two} for a discussion, in the pure YM case, on how this property is related to the positivity of the integration measure under the functional integral. Its extension in the presence of dynamical quarks is discussed below.} of $\smash{\tilde\Gamma[\bar A]\equiv\Gamma[\bar A,\langle a\rangle=0]}$, where $\Gamma[\bar A,\langle a\rangle]$ is the effective action for $\langle a\rangle$ in the presence of $\bar A$ \cite{Braun:2007bx,reinosa2016two}.  It is easily shown that $\tilde\Gamma[\bar A]$ is invariant under center transformations. Therefore, center symmetry is manifest at every step and the deconfinement transition can be monitored by studying when the minima of $\tilde\Gamma[\bar A]$ depart from their center-symmetric (confining) values, see Refs.~\cite{reinosa2016two,Herbst:2015ona} for more details. 

The minima are usually sought for in the subspace of configurations that comply explicitly with the symmetries of the system at finite temperature. In particular, one restricts to temporal and homogenous backgrounds $\smash{\bar A_\mu(\tau,{\bf x})=\bar A_0\delta_{\mu 0}}$ and the functional $\tilde\Gamma[\bar A]$ reduces to an effective potential $V(\bar A_0)$ for the constant matrix field $\bar A_0$. Without loss of generality, one can choose this matrix to lie in the Cartan subalgebra:
\beq\label{eq:bg0}
\beta g\bar A_0=r_j\, t^j\,,
\eeq
where the $r_j$'s are dimensionless components, the $t^j$'s span the Cartan subalgebra and a summation over $j$ is implied. 

The potential becomes then a function $V(r)$ of the vector $r$ that can be evaluated within a loop expansion at high temperatures \cite{Dumitru:2013xna}. The same expansion is not expected to be valid, however, in the low temperature phase, in particular due to the presence of Gribov copies that make the LDW gauge-fixing ambiguous in the infrared. The perturbative expansion has recently been modified to try to account for the effect of the Gribov copies. To this purpose, a mass term
\begin{equation}
  \label{eq:mass}
  S_m=\int_x \frac 12 m^2 a_\mu^a a_\mu^a\,,
\end{equation}
has been added to the action (\ref{eq:fp}), see \cite{Reinosa:2014ooa}. The background field effective potential in the presence of such a mass term has been computed to one- and two-loop orders for SU($N$) Yang-Mills theories \cite{Reinosa:2014ooa,Reinosa:2014zta,reinosa2016two}. It leads to a confinement-deconfinement transition with the expected order depending on the value of $N$.

\subsection{Dynamical quarks}\label{sec:add_quarks}
Including dynamical quarks is straightforward, at least in the heavy-quark limit, where chiral symmetry breaking is not an issue \cite{Lo:2014vba,Fukushima:2017csk}. The one-loop contribution to the background field potential in the present framework can be found, for instance, in Ref.~\cite{reinosa2015perturbative}. Together with the corresponding one-loop contribution from the gauge sector in the presence of the mass term (\ref{eq:mass}), this correctly describes most of the qualitative and quantitative aspects of the phase diagram at nonzero $T$ and $\mu$, known from lattice simulations \cite{Fromm:2011qi}. The purpose of this paper is to compute the two-loop correction from the quark sector, add it to the corresponding correction from the gauge sector computed in Ref.~\cite{reinosa2016two}, and check the convergence properties of the expansion. 

In Refs.~\cite{reinosa2015perturbative,Reinosa:2016xaj}, it was also pointed out that, depending on the context, some of the components of the background $\bar A_0$ need to be continued from the real to the imaginary axis, as we now recall.\footnote{This discussion is similar to the one in terms of the Polyakov loop variables $\ell$ and $\bar\ell$ (to be introduced below) see Ref.~\cite{Dumitru:2005ng}. In the context of the Landau-deWitt gauge with self-consistent backgrounds, it was  understood only recently \cite{reinosa2015perturbative} that some of the background components need to be continued in the case of a real chemical potential, as required by the very condition of existence of a self-consistent background in the presence of a complex integration measure under the functional integral. The need for complex background components was also pointed out in the context of the saddle point approximation in Refs.~\cite{Nishimura:2014rxa,Nishimura:2014kla}.} In the SU(3) case, the decomposition (\ref{eq:bg0}) reads
\beq\label{eq:bg}
\beta g\bar A_0=r_3\frac{\lambda_3}{2}+r_8\frac{\lambda_8}{2}\,,
\eeq
with $\lambda_3$ and $\lambda_8$ the diagonal Gell-Mann matrices. In the presence of an imaginary chemical potential (including the case of vanishing chemical potential as well as the pure Yang-Mills case), one can argue that the background effective potential is a real function when $\smash{r\equiv(r_3,r_8)}$ is taken in the plane $\mathds{R}\times\mathds{R}$ and that the self-consistent backgrounds correspond to the absolute minima of the effective potential in that plane. In fact, the effective potential being invariant under (periodic) gauge transformations that preserve the form of the background (\ref{eq:bg}),\footnote{These include particular global color rotations, known as Weyl transformations.} this plane is divided into physically equivalent cells, referred to as {\it Weyl chambers}. In practice, it is thus enough to restrict to one of these chambers, for instance the equilateral triangle of edges $(0,0)$ and $2\pi(1,\pm 1/\sqrt{3})$, which we call the {\it fundamental} Weyl chamber in what follows. In this chamber, the confining point is located at $\smash{r=(4\pi/3,0)}$. Moreover, the median $r_8=0$ corresponds to charge conjugation invariant states. It follows that, as long as the chemical potential is zero, one can restrict to this axis for the purpose of determining the physical point. When an imaginary chemical potential is introduced, the physical point moves away from the axis $r_8=0$ in the Weyl chamber.

In contrast, for a real chemical potential, it was argued in Ref.~\cite{reinosa2015perturbative} that the effective potential needs to be considered over the space $(r_3,r_8)\in\mathds{R}\times i\mathds{R}$, where it remains real (whereas it is not anymore over $\mathds{R}\times\mathds{R}$). The price to pay is, however, that the physical point corresponds in this case to a saddle-point and it is not clear which one to choose when multiple saddle-points are present.\footnote{This is in contrast to the case of an imaginary chemical potential, where standard arguments based on the positivity of the fermion determinant dictate that the physical point corresponds to the absolute minima of the background effective potential.} In Ref.~\cite{reinosa2015perturbative}, this was interpreted as a remnant of the sign problem in continuum approaches. It was also proposed (based on the limit of small chemical potential) that the appropriate criterion might be to choose the deepest saddle-point. We shall use this criterion in this work as well. For a similar discussion in terms of the Polyakov loops, see Ref.~\cite{Fukushima:2006uv} and, for a connection between the two pictures, see Ref.~\cite{Reinosa:2016xaj}.

\subsection{Polyakov loops}
The standard, gauge invariant order parameters for the deconfinement transition are the averages of the traced Polyakov loops in the fundamental representations ${\bf 3}$ and ${\bf \bar 3}$. They are defined as
\begin{align}
\label{eq_popoldef}
\ell&\equiv\frac 13\tr \left\langle P \exp\left(i g\int_0^\beta \!d\tau A_0^at^a\right)\right\rangle,\\
\label{eq_popoldefbar}
\bar \ell&\equiv\frac 13\tr \left\langle \bar P \exp\left(-i g\int_0^\beta \!d\tau A_0^at^a\right) \right\rangle,
\end{align}
where $P$ and $\bar P$ denote path ordering and anti path ordering, respectively. These expectation values can be computed perturbatively in terms of the physical values of the backgrounds $r_3$ and $r_8$. For instance, at leading-order, one has the well-known expression
\beq\label{eq:pl1}
\ell=\frac{e^{-i\frac{r_8}{\sqrt{3}}}+2\cos(r_3/2)\,e^{i\frac{r_8}{2\sqrt{3}}}}{3}+{\cal O}\left(g^2\right).
\eeq
The next-to-leading order correction to $\ell$ in the present framework is given by the pure YM expressions derived in Ref.~\cite{reinosa2016two}; see Eq.~(87) of that reference. This is because, at next-to-leading order, dynamical quarks only enter through the actual values of $r_3$ and $r_8$ at which the expressions \eqn{eq_popoldef} and \eqn{eq_popoldefbar} must be evaluated.\footnote{Explicit quark loops would appear only at higher orders.} However, in order to derive a similar expression for $\bar\ell$, one must pay attention to the above remarks concerning the different spaces the background component $r_8$ needs to be varied over depending on the (real or imaginary) value of the chemical potential.

For an imaginary chemical potential, both $r_3$ and $r_8$ are real and one has $\bar\ell=\ell^*$ \cite{reinosa2015perturbative}, in line with the general discussion of Ref.~\cite{Dumitru:2005ng}. In particular, at leading order, one has
\beq
\label{LOexprlbar}
\bar\ell=\frac{e^{i\frac{r_8}{\sqrt{3}}}+2\cos(r_3/2)\,e^{-i\frac{r_8}{2\sqrt{3}}}}{3}+{\cal O}\left(g^2\right).
\eeq
Similarly, the next-to-leading correction is simply the complex conjugate of Eq.~(87) of Ref.~\cite{reinosa2016two}. In the case of a real chemical potential, one analytically continues the so-obtained expression for $\bar\ell$ from $r_8\in\mathds{R}$ to $r_8\in i\mathds{R}$. It is easily checked that this continuation yields $\bar\ell\in\mathds{R}$ for real $\mu$, as expected from general arguments \cite{reinosa2015perturbative,Dumitru:2005ng}. This is obvious on the leading-order expression \eqn{LOexprlbar} and can be explicitly checked at next-to-leading order as well. It is also not very difficult to convince oneself that this recipe of analytic continuation coincides with the direct computation of $\bar\ell$, e.g., along the lines of Ref.~\cite{reinosa2016two}, in a background $(r_3,r_8)\in\mathds{R}\times i\mathds{R}$.

\section{Two-loop corrections}
\label{sec:calculation}
\begin{center}
\begin{figure}[t]  
\epsfig{file=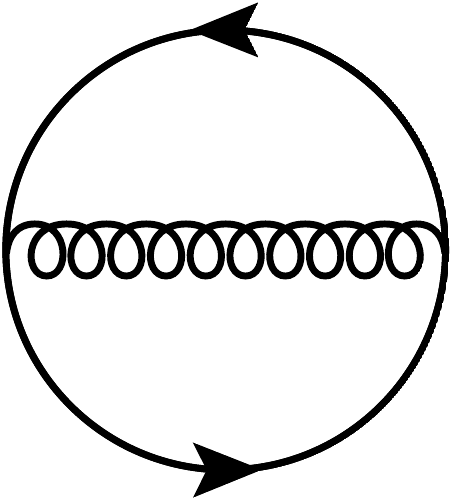,width=2.7cm}
 \caption{Two-loop contribution to the background field effective potential coming from dynamical quarks.}\label{fig:Quarksunset}
\end{figure}
\end{center}

The pure-glue one-loop contribution to the potential, which we denote by $V_g(r,T)$ in what follows, can be found in Refs.~\cite{reinosa2016two,Reinosa:2014ooa}. The two-loop contribution $V_g^{(2)}(r,T)$ has been computed in Ref.~\cite{Reinosa:2014zta} for the SU(2) case and in Ref.~\cite{reinosa2016two} for SU($N$),\footnote{In fact, for any compact Lie group with a simple Lie algebra.} with $N\geq 3$. The one-loop quark contribution, denoted $V_q(r,T,\mu)$,  is a well-known result \cite{Lo:2014vba,Fukushima:2017csk,reinosa2015perturbative}. For definiteness, we refer to Eq.~(58) of Ref.~\cite{reinosa2015perturbative}. In the following, we focus on the two-loop contribution from the quark sector, denoted $V_q^{(2)}(r,T,\mu)$ and given by the quark sunset diagram shown in Fig.~\ref{fig:Quarksunset}.

\subsection{Feynman Rules}
\begin{center}
\begin{figure}[h]  
\epsfig{file=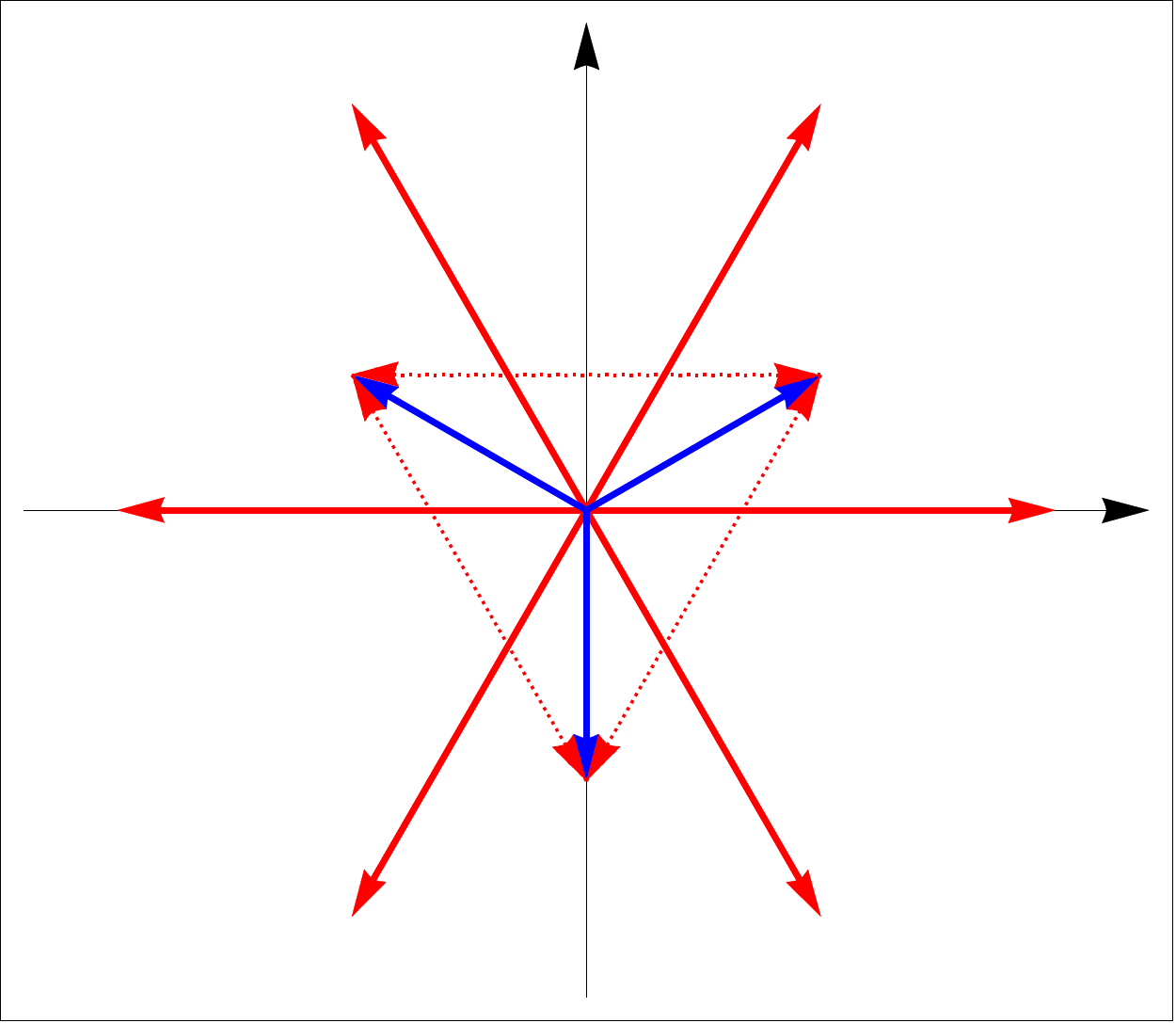,width=6.1cm}
 \caption{The six roots of the su(3) algebra (red) and the three weights of the fundamental representation ${\bf 3}$ (blue). The dotted lines represent the roots that have been translated to illustrate the relation between the roots and the weights. The roots always come in pairs $\pm\alpha$ and each given root connects two particular weights. This rule is nothing but color conservation at the quark-gluon vertex.}\label{fig:root}
\end{figure}
\end{center}

In the presence of the background (\ref{eq:bg0}), the Feynman rules take a simple form provided one decomposes the various fields along bases that diagonalize the action of the background covariant derivatives $\bar D$ and $\bar{\cal D}$ in the adjoint and fundamental representations respectively. For the former, one introduces a Cartan-Weyl basis $\{t^\kappa\}$ in the SU($N$) Lie algebra, where $\kappa$ can take two types of values: either $\kappa=0^{(j)}$ is a ``zero'' in which case the corresponding $t^{0^{(j)}}$'s (which are nothing but the $t^j$'s introduced earlier) span the Cartan subalgebra, or $\kappa=\alpha$ is a root in which case the corresponding $t^\alpha$'s simultaneously diagonalize the action of the elements of the Cartan subalgebra in the adjoint representation:
\beq
[t^{0^{(j)}},t^\alpha]=\alpha_j t^\alpha\,.
\eeq
The roots $\alpha$ are vectors with as many (real) components as there are elements in the Cartan subalgebra. It is thus common to represent them in the space $\mathds{R}^{d_C}$, with $d_C$ the dimension of the Cartan subalgebra; they form what is called the root diagram of the algebra. For SU(3), we have two zeros $0^{(3)}$ and $0^{(8)}$ to which correspond $\smash{t^{0^{(3)}}=\lambda_3/2}$ and $\smash{t^{0^{(8)}}=\lambda_8/2}$. We also have six roots $\pm (1,0)$, $\pm(1,\sqrt{3})/2$ and $\pm (1,-\sqrt{3})/2$. The corresponding root diagram is shown in Fig.~\ref{fig:root}.\\

Decomposing the gluon field along a Cartan-Weyl basis, the corresponding gluon propagator is obtained as
\beq
\langle a_\mu^{-\kappa}(Q')\,a_\nu^{\kappa}(Q)\rangle =(2\pi)^d\delta^{(d)}(Q+Q') \frac{P^\perp_{\mu \nu}(Q_\kappa )}{Q_\kappa^2+m^2}\,,
\eeq
where we have introduced the generalized momentum $\smash{Q^\mu_\kappa\equiv Q^\mu+\delta^{\mu0}\,T\,r\cdot\kappa}$, with $\smash{Q_\mu=(\omega_n,\vec{q})}$, $\smash{\omega_n=2\pi nT}$ a bosonic Matsubara frequency and $r\cdot\kappa\equiv r_j\kappa^j$. One can similarly obtain the ghost propagator as well as the three- and four-gluon interaction vertices \cite{reinosa2016two} but they will not be needed for the present work. For later use, we note, that $Q^\kappa=-(-Q)^{-\kappa}$.\\

In the quark sector, we introduce vectors that diagonalize simultaneously the action of the $t^{0^{(j)}}$'s in the fundamental representation. We shall denote these vectors using the ket notation $|\rho\rangle$. We have then
\beq
t^{0^{(j)}}|\rho\rangle=\rho_j |\rho\rangle\,,
\eeq
where the $\rho$'s are called the weights of the (fundamental) representation. They are vectors with the same number of (real) components $\rho_j$ as the roots and can therefore be represented on the same diagram.\footnote{The vectors $|\rho\rangle$ and $\rho$ have of course a different meaning. The former are vectors in the space over which the representation acts. In contrast, the latter are just a convenient way to collect the various eigenvalues arising from the simultaneous diagonalization of the generators $t^{0^{(j)}}$.} The fundamental representation ${\bf 3}$ has three nondegenerate weights $(1,1/\sqrt{3})/2$, $(-1,1/\sqrt{3})/2$ and $(0,-1/\sqrt{3})$; see Fig.~\ref{fig:root}.

We mention that, the $t^{0^{(j)}}$ being hermitian, the $|\rho\rangle$'s can be chosen such that $\langle\rho|\sigma\rangle=\delta_{\rho\sigma}$. Now, by decomposing the Grassmann fields $\psib_f$ and $\psi_f$ as $\psib_f=\sum_{\rho}\psib_{f\rho} \bra{\rho}$ and $\psi_f=\sum_{\rho}\psi_{f\rho} \ket{\rho}$, where $\psi_{f\rho}$ is a bispinor without color structure, one obtains the quark propagator as
\beq
\langle \psi_{f\rho}(P')\psib_{f\rho}(P)\rangle\!=\!(2\pi)^d\delta^{(d)}(P+P') \frac{i\slashed{P}_\rho{+}M_f}{P_\rho^2+M_f^2}\,,
\eeq
where we have introduced\footnote{In what follows, we use the greek letters $\mu$ and $\nu$ to denote Lorentz indices, $\rho$ and $\sigma$ to denote the weights of the fundamental representation and $\kappa$ to denote the zeros and roots of the algebra. Beware that $\mu$ also refers to the chemical potential when it is not a sub/superscript. For simplicity, we use the same notation for bosonic and fermionic four-momenta. To avoid confusion though, we shall reserve the letter $Q$ to bosonic momenta and the letters $P$ and $L$ to fermionic ones.} $\smash{P^\nu_\rho\equiv  P^\nu+\delta^{\nu0}\,(T\,r\cdot\rho +i\mu)}$, with $ P^\nu\equiv(\hat\omega_n,\vec{p})$, $\hat\omega_n=2\pi(n+1/2)T$ a fermionic Matsubara frequency and $r\cdot\rho\equiv r_j\rho^j$. Note that the fermionic frequency $P^0_\rho$ is shifted not only by $T\,r_j\rho^j$ (similar to the shift of the bosonic frequencies by $T\,r_j\kappa^j$) but also by $i\mu$. This is the well-known fact that the chemical potential can be seen as an imaginary background and vice versa.\footnote{More precisely, the ``baryonic'' chemical potential considered here plays the role of an imaginary temporal abelian background. Reversely, the components of the background $\bar A_0$ along the Cartan directions correspond to imaginary chemical potentials for the color charges along the same directions.} As a consequence, for nonzero chemical potential, one must pay attention to the fact that $ P^\rho \neq -(- P)^{-\rho},$ in contrast to the similar formula given above in the gluon sector. For this identity to be true here, one needs to change simultaneously the sign of the chemical potential, $\mu \rightarrow -\mu$.

Finally, the quark-gluon vertex arises from the interaction term 
\beq
i\psib_f a^\kappa t^\kappa \psi_f=\sum_{\rho,\sigma}i\psib_{f\sigma} a^\kappa \langle\sigma|t^\kappa|\rho\rangle \psi_{f\rho}\,,
\eeq
where we have included the minus sign coming from $-S$. The corresponding Feynman rule is then
 \beq
 i g\, \delta(P+Q+P')\,\gamma_\mu t^\kappa_{\sigma\rho}\,,
 \eeq 
where $\smash{t^\kappa_{\sigma\rho}\equiv\langle\sigma|t^\kappa|\rho\rangle}$. We mention that 
 \begin{eqnarray}
t^{0^{(j)}}t^\kappa\, \ket{\rho} &=& [t^{0^{(j)}},t^\kappa]\,\ket{\rho} +t^\kappa t^{0^{(j)}}\,\ket{\rho}  \nonumber\\
&=& (\kappa + \rho)_j\, t^\kappa \, \ket{\rho}\,,
\end{eqnarray}
implying that $t^\kappa|\rho\rangle$ is either $0$ or collinear to $|\rho+\kappa\rangle$. In particular, the nonvanishing elements of $t^\kappa_{\sigma\rho}$ satisfy the color conservation rule $\sigma=\rho+\kappa$. The relation between the roots and the weights of the su(3) algebra is illustrated in the root diagram of Fig.~\ref{fig:root}.

\subsection{Quark sunset contribution}
Let us now use the Feynman rules listed above in order to evaluate the (two-loop) quark sunset diagram contribution to the background field potential. We shall regularize the UV divergences by working in $d=4-2\epsilon$ dimensions. Moreover, we introduce the following notations for the bosonic and fermionic Matsubara sum-integrals:
\bea\label{eq:eq}
\int_Q f(Q) & \equiv & T\sum_n\mu_r^{2\epsilon}\int\frac{d^{d-1}q}{(2\pi)^{d-1}} f(\omega_n,q),\\
\int_{\hat P} f(P) & \equiv & T\sum_n\mu_r^{2\epsilon}\int\frac{d^{d-1}p}{(2\pi)^{d-1}} f(\hat\omega_n,p),
\eea
with $\mu_r$ an arbitrary scale. With these conventions, the quark sunset diagram contribution to the background field effective potential reads
\begin{eqnarray}\label{Quarksunset}
&& V_q^{(2)}(r,T,\mu)=\nonumber\\
&&\hspace{0.05cm}-\,\frac{g^2}{2}  \sum_f \sum_{\sigma \rho \kappa} {\cal D}_{\sigma,\rho \kappa}\int_{\hat P} \int_Q G_{m}(Q^\kappa)G_{M_f}(P^\rho) G_{M_f}(L^\sigma) \nonumber \\ 
&& \hspace{0.7cm}\times\,P^\perp_{\mu\nu}(Q^\kappa) \, {\rm tr}\, \big[ \gamma_\mu (i\slashed{P}^\rho+M_f) \gamma_\nu (i\slashed{L}^\sigma+M_f)    \big],      
\end{eqnarray}
where we have introduced the notation
\beq
G_{m}(X)=\frac{1}{X^2+m^2}\,,
\eeq
and where $L\equiv P+Q$, so both $P^0$ and $L^0$ are fermionic Matsubara frequencies. We have also defined ${\cal D}_{\sigma,\rho \kappa}\equiv t^{\kappa}_{\sigma\rho}t^{-\kappa}_{\rho\sigma}$. The properties of the tensor $t^\kappa_{\rho\sigma}$ imply that the only nonvanishing elements of ${\cal D}_{\sigma,\rho\kappa}$ are such that $\smash{\sigma=\rho+\kappa}$. This conservation law follows from the invariance of the LDW gauge-fixed action under color rotations that leave the background (\ref{eq:bg0}) invariant, that is, color rotations with generators in the Cartan subalgebra. It implies that the standard conservation of momentum, $\smash{L=P+Q}$, extends to the generalized momenta in the form $\smash{L^\sigma=P^\rho+Q^\kappa}$. Finally, we note the obvious property ${\cal D}_{\sigma,\rho\kappa}={\cal D}_{\rho,\sigma(-\kappa)}$.

After dealing with the trace structure, one can further reduce the above expression in terms of scalar (bosonic or fermionic) sum-integrals. This is detailed in Appendix \ref{Reduction}.
We obtain
\begin{eqnarray}\label{ScalarInts}
V_q^{(2)}(r,T,\mu)&=&-\,\frac{g^2}{4}{\rm tr}\mathds{1}   \sum_f \sum_{\sigma \rho \kappa} {\cal D}_{\sigma , \rho \kappa } \nn
&\times&\Bigg\{(d-2)  \bigg[ J^\kappa_m\left(J^\rho_{M_f}  + J^\sigma_{M_f}\right) -J^\rho_{M_f}J^\sigma_{M_f}  \bigg] \nn
&&+\,\frac{2}{m^2}\bigg[ \left( \tilde{J}^\kappa_0-\tilde{J}^\kappa_m \right)\left(\tilde{J}^\rho_{M_f} -\tilde{J}^\sigma_{M_f}\right)\bigg]\nn
&&+\bigg[(d-2)m^2+4M_f^2\bigg]  S_{mM_fM_f}^{\kappa\rho\sigma} \Bigg\},
\end{eqnarray}
where we note that the prefactor ${\rm tr}\mathds{1}$ is not uniquely defined in dimensional regularization. The only constraint is that it should approach $4$ in the limit $\epsilon\to 0$. We shall check that our final result (for the medium dependent part of the background field effective potential) does not depend on the particular way this limit is approached if the same definition for ${\rm tr}\mathds{1}$ is used consistently everywhere. 

We have introduced the bosonic scalar tadpoles 
\begin{eqnarray} \label{bosonicTads}
		J^\kappa_m\equiv\int_Q G_m(Q^\kappa) \quad {\rm and} \quad \tilde{J}^\kappa_m\equiv\int_Q Q^\kappa_0 G_m(Q^\kappa)\, , \\ \nonumber 
	\end{eqnarray}
	the fermionic scalar tadpoles\footnote{To avoid complicated notations, we use the same letters for bosonic and fermionic tadpole integrals and we distinguish them by their color labels: a zero or a root ($\kappa$) for the former and a weight ($\rho$) for the latter.}
	\begin{eqnarray} \label{fermionicTads}
		J^\rho_{M}\equiv\int_{\hat P} G_M(P^\rho)\quad {\rm and} \quad \tilde{J}^\rho_M\equiv\int_{\hat P} P^\rho_0 G_M(P^\rho)\,,
	\end{eqnarray}
	as well as the fermion--boson scalar sunset
	\begin{eqnarray}\label{scalarSunset}
		S_{mMM}^{\kappa\rho\sigma}\equiv\int_{\hat P}\int_Q   G_m(Q^\kappa)G_M(P^\rho)G_M(L^\sigma)\,.
	\end{eqnarray}
Performing the bosonic and fermionic Matsubara sums in these one- and two-loop sum-integrals produces various contributions involving up to two Bose--Einstein and/or Fermi-Dirac distribution functions; see Appendix \ref{ThermalFactors}. It is useful to split these contributions according to the number ---zero ($0n$), one ($1n$), or two ($2n$)--- of such thermal factors. For instance, we split the various tadpole integrals introduced before as 
	\beq
	  J = J(0n)+J(1n)
	  \eeq
	  and the sunset sum-integral as
	  \begin{eqnarray}
		S_{mMM}^{\kappa\rho\sigma} =\, S_{mMM}^{\kappa\rho\sigma}(0n)+S_{mMM}^{\kappa\rho\sigma}(1n)+S_{mMM}^{\kappa\rho\sigma}(2n).
	\end{eqnarray}
	
	The vacuum contribution to the background field potential, with no thermal factors, depends neither on the background nor on the temperature or the chemical potential and, therefore, does not enter the determination of the physical values of the background or the in-medium dependence of thermodynamical observables. We discard it systematically in what follows.

When collecting terms with one thermal factor, there appear sums of the form $\sum_{\rho\kappa} {\cal D}_{\sigma , \rho \kappa }$ and $\sum_{\sigma\rho}{\cal D}_{\sigma , \rho \kappa }$. The former is nothing but the Casimir $C_F$ of the fundamental representation, which equals $4/3$ in the SU(3) case, and the latter is the normalization $T_F$ for the generators in the fundamental representation, chosen as $1/2$. After the calculations described in Appendix~\ref{ThermalFactors}, we arrive at the following expression for the quark sunset contribution to the background field effective potential (from which we have removed the pure vacuum part):
\begin{widetext}
\begin{eqnarray}\label{eq:res}
 V_q^{(2)}(r,T,\mu) & = & -g^2{\rm tr}\mathds{1}  \sum_f\Bigg\{ \frac{1}{4}\Big[ (d-2) J_{M_f}(0n) +\frac{1}{2}\left[(d-2)m^2+4M_f^2\right]I_{M_f M_f}(0n)\Big] \sum_{\kappa} J_m^{\kappa}(1n)\nonumber\\
& & \hspace{1.0cm}+ \frac{C_F}{2}\Big[ (d-2)\left[J_m(0n)-J_{M_f}(0n) \right]+ \left[(d-2)m^2+4M_f^2\right] I_{M_f m}(0n) \Big]\sum_{\rho} J_{M_f}^{\rho}(1n) \Bigg\}\nonumber\\
& - & \frac{g^2}{2}{\rm tr}\mathds{1}\sum_f \sum_{\sigma \rho \kappa} {\cal D}_{\sigma , \rho \kappa } \Bigg\{ \left[J_{M_f}^{\rho}(1n)+J_{M_f}^{\sigma}(1n)\right]J_{m}^{\kappa}(1n)-J_{M_f}^{\rho}(1n)J_{M_f}^{\sigma}(1n)\nonumber\\
& & \hspace{1.0cm}+\frac{1}{m^2}\left[\tilde{J}^\kappa_0(1n)-\tilde{J}^\kappa_m(1n) \right]\left[\tilde{J}^\rho_{M_f}(1n)-\tilde{J}^\sigma_{M_f}(1n)\right]+(m^2+2M_f^2) S_{mM_fM_f}^{\kappa\rho\sigma}(2n)\Bigg\},
\end{eqnarray}
\end{widetext}
where the expressions for $J_m(0n)$, $J_M(0n)$, $I_{Mm}(0n)$, $I_{MM}(0n)$, $J_m^\kappa(1n)$, $\tilde J_m^\kappa(1n)$, $J^\rho_M(1n)$, $\tilde J^\rho_M(1n)$ and $S_{mMM}^{\kappa\rho\sigma}(2n)$ are given in Appendix \ref{ThermalFactors}.

The ($0n$) contributions in the first two lines of Eq.~(\ref{eq:res}) are UV divergent. The renormalization of the background field potential will be dealt with in the next section. In contrast, the last two lines of Eq.~(\ref{eq:res}) are UV finite and one can take the limit $d\to 4$, including the replacement ${\rm tr}\mathds{1}\to 4$. Moreover, this finite contribution can be further simplified as follows. The sum over $\kappa$ is a sum over the zeros $0^{(j)}$ and the roots $\alpha$. When $\kappa$ is a zero $0^{(j)}$, we can use that $\sigma=\rho$ and ${\cal D}_{\rho,\rho 0^{(j)}}=\rho_j^2$, as well as the fact that $J_m^{0^{(j)}}(1n)$, $\tilde J_m^{0^{(j)}}(1n)$ and $S_{mMM}^{0^{(j)}\rho\rho}(2n)$ do not depend on $j$, in order to rewrite the sum over the zeros as
\begin{align}
 -2g^2\sum_{f}\sum_{\rho}\rho^2 &\Big\{ J_{M_f}^{\rho}(1n)\left[2J_{m}^0(1n)-J_{M_f}^{\rho}(1n)\right]\nonumber\\
&+\left(m^2+2M_f^2\right) S_{mM_fM_f}^{0\rho\rho}(2n)\Big\},
\end{align}
with $\rho^2=1/3$ in the SU(3) case. As for the sum over the roots $\alpha$, no such simplifications occurs. However, we can make use of ${\cal D}_{\sigma,\rho\alpha}={\cal D}_{\rho,\sigma(-\alpha)}$, $J^\alpha(1n)=J^{-\alpha}(1n)$, $\tilde J^\alpha(1n)=-\tilde J^{-\alpha}(1n)$, and $S^{\alpha\rho\sigma}_{mMM}(2n)=S^{(-\alpha)\sigma\rho}_{mMM}(2n)$ to keep in the sum only one out of the two contributions $(\sigma,\rho,\alpha)$ and $(\rho,\sigma,-\alpha)$. This is only valid if we use a summand that has the above mentioned symmetry, which is the case for the expression in Eq.~(\ref{eq:res}). For SU(3), there are three such independent contributions (see Fig.~\ref{fig:root}) and for each of them ${\cal D}_{\sigma,\rho\alpha}=1/2$.

\subsection{Renormalization}
Let us now deal with the first two lines of Eq.~(\ref{eq:res}), which are UV divergent. A straightforward calculation using the expressions for $J_M(0n)$ and $I_{MM}(0n)$ given in Appendix \ref{ThermalFactors}, shows that the prefactor in front of $\sum_\kappa J_m^\kappa(1n)$ is\footnote{We display here the case $M_f>2m$ but the other case can be treated similarly using the expressions in App.~\ref{ThermalFactors}.}
\begin{align}\label{eq:pref1}
  -\frac{g^2{\rm tr}\,\mathds{1}}{64\pi^2}\sum_f\Bigg\{&m^2\left[\frac{1}{\epsilon}+\ln\frac{\bar\mu^2}{M^2_f}\right]+m^2+4M^2_f\nonumber\\
& -\,2\left(m^2+2M^2_f\right){\cal T}_f\,\arctan\left({\cal T}_f^{\,-1}\right)\Bigg\},
\end{align}
with ${\cal T}_f\equiv\sqrt{\frac{4M^2_f}{m^2}-1}$ and $\bar\mu^2=4\pi \mu_r^2 e^{-\gamma}$, with $\gamma$ the Euler-MacLaurin constant. Similarly, the prefactor in front of $\sum_\rho J_{M_f}^\rho(1n)$ reads
\begin{align}\label{eq:pref2}
\frac{g^2C_F{\rm tr}\,\mathds{1}}{16\pi^2}\Bigg\{&m^2-3M^2_f\left[\frac{1}{\epsilon}+\ln\frac{\bar\mu^2}{M^2_f}\right]-\left(m^2+2M^2_f\right)\nn
&\times\left[2-\frac{m^2}{2M^2_f}\ln \frac{m^2}{M^2_f}-\frac{m^2}{M^2_f}{\cal T}_f\,\arctan\left({\cal T}_f\right)\right]\Bigg\}.\nonumber\\
\end{align}
The corresponding divergences are absorbed by two one-loop diagrams involving counterterms. There is a quark loop
\begin{eqnarray}
& & -\sum_f\int_{\hat P}{\rm tr}(-\delta Z_{\psi_f} i P\!\!\!\!\slash_\rho+\delta M_f)(iP\!\!\!\!\slash_\rho+M_f)G_{M_f}(P_\rho)\nonumber\\
& & = -\sum_f\int_{\hat P} (\delta Z_{\psi_f} P^2_\rho+M_f\delta M_f)G_{M_f}(P_\rho)\times {\rm tr}\mathds{1}\nonumber\\
& & = -\sum_f M_f(\delta M_f-M_f\delta Z_{\psi_f})\sum_\rho J^\rho_{M_f}\times {\rm tr}\mathds{1}\,,
\end{eqnarray}
as well as a gluon loop
\beq
\frac{d-1}{2}(\delta m^2_q-m^2\delta Z_{A,q})\sum_\kappa J^\kappa_m\,.
\eeq
In the latter, we have only kept the contribution to the counterterms that originates from the quarks, hence the subscript $q$. The pure glue contributions to $\delta Z_A$ and $\delta m^2$ have already been taken into account in the renormalization of the pure glue potential at two-loop order \cite{reinosa2016two}.

The quark contribution to the background field potential \eqn{eq:res} is made finite if the divergent parts of the counterterms satisfy 
\beq
\left[\delta M_f-M_f\delta Z_{\psi_f}\right]_{\rm div}=-\frac{3g^2C_FM_f}{16\pi^2\epsilon}\,,
\eeq
and
\beq
\left[\delta m^2_q-m^2\delta Z_{A,q}\right]_{\rm div}=\frac{g^2m^2}{24\pi^2\epsilon}N_f\,.
\eeq
As expected, these conditions agree with those obtained in Ref.~\cite{Pelaez:2014mxa} for the renormalization of the quark contribution to the vacuum gluon and quark self-energy at one-loop order, respectively.\footnote{At $T=\mu=0$, the LDW gauge reduces to the Landau gauge considered in that reference. Note also that the authors use a different sign convention for the self-energy.} In fact, it is not difficult to check that the prefactor (\ref{eq:pref1}) in front of $\sum_\kappa J_m^\kappa(1n)$ can be written
\beq\label{eq:cont}
\frac{d-1}{2}\Pi^\perp_{q}(Q^2\to-m^2)\,,
\eeq
where $\Pi^\perp_{q}(Q^2\to-m^2)$ is defined as the real part of the quark-loop contribution to the transverse gluon self-energy, computed in the vacuum with the same convention for ${\rm tr}\,\mathds{1}$ and continued from $Q^2>0$ to $Q^2=-m^2-i0^+$. This can be checked by a direct calculation using the formulae in Ref.~\cite{Pelaez:2014mxa}. The contribution (\ref{eq:cont}) can be conveniently combined with the gluon counterterm diagram and yields
\beq\label{eq:fghj}
\frac{d-1}{2}\left[\delta m^2_q-m^2\delta Z_{A,q}+\Pi^\perp_{q}(Q^2\to -m^2)\right].
\eeq
Similarly, the prefactor in front of $\sum_\rho J_{M_f}^\rho(1n)$ including the quark counterterm contribution rewrites
\beq\label{eq:fghj2}
{\rm tr}\,\mathds{1}\,M_f^2\left\{\delta Z_{\psi_f}-\frac{\delta M_f}{M_f}+\left[A_f-B_{f}\right](P^2\to -M^2_f)\right\},
\eeq
where we have decomposed the one-loop vacuum quark self-energy as $\Sigma_f(P)=-iA_{f}(P^2)\,\slash{\!\!\!\!P}+M_fB_{f}(P^2)$. Not only do these expressions make it transparent why the needed divergent contributions to the counterterms are those given by the renormalization of the one-loop gluon and quark self-energies, but they also allow us to express the effect of various renormalization schemes in a convenient way. 

Consider for instance the generic set of renormalization conditions
\begin{eqnarray}
0 & = & \delta m^2_q+\Pi^\perp_{q}(0)\,,\\
0 & = & \delta Z_{A,q}+\frac{\delta m^2_q+\Pi^\perp_{q}(\mu_r^2)}{\mu_r^2}\,,\\
0 & = & \delta Z_{\psi_f}+A_{f}(\hat\mu_{f}^2)\,,\\
0 & = & \delta M_f+M_fB_{f}(\mu_{f}^2)\,,
\end{eqnarray}
which coincides with the one used in the calculation of the two-loop pure glue potential in Ref.~\cite{reinosa2016two} but leaves some freedom in the quark sector through the arbitrary scales $\hat\mu_{f}$ and $\mu_{f}$. In the gluon sector one finds in particular that $\delta m^2_q=0$ and the one-thermal-factor-contribution in Eq.~(\ref{eq:res}) rewrites
\begin{align}
& \frac{3}{2}m^2\left[\frac{\Pi^\perp_{q}(\mu^2)}{\mu^2}-\frac{\Pi^\perp_{q}(Q^2\to-m^2)}{-m^2}\right]\sum_\kappa J^\kappa_m(1n)\nonumber\\
& +\,{\rm tr}\,\mathds{1}\sum_fM^2_f\,\bigg\{\left[B_{f}(\mu^2_{f})-B_{f}(P^2\to-M^2_f)\right]\nonumber\\
\label{eq:fgegf}& -\left[A_{f}(\hat\mu^2_{f})-A_{f}(P^2\to-M^2_f)\right]\bigg\}\sum_\rho J^\rho_{M_f}(1n).
\end{align}
Various remarks are in order here. First, in this form, the cancellation of divergences is explicit. In particular, the result of the last two lines is independent of the way ${\rm tr}\,\mathds{1}$ is sent to $4$. The same is true for the first line because both terms in the (UV finite) difference are proportional to ${\rm tr}\,\mathds{1}$. Second, the bracket of the first line yields the only polynomial contribution ($\sim 1/M_f^2$) in the large quark mass limit,\footnote{In fact, each integral in the bracket behaves as $\ln M_f$ in this limit, however, with a coefficient independent of the momentum, so that the difference is suppressed.} the other contributions being exponentially suppressed. Such a polynomial behavior is an effect of the gluon mass introduced in the present model\footnote{The prefactor in front of $\sum_\kappa J^\kappa_m(1n)$ vanishes for $m^2=0$ because, in this case, BRST symmetry, applied both to the QCD and YM cases, imposes $\Pi^\perp_{q}(Q^2=0)=0$.}  and one may wonder whether such effect could be tested on the lattice by studying the approach to the quenched (pure YM) limit. We note, however, that there exists a particular scheme where the polynomial approach to YM is replaced by an exponential behavior. This corresponds to choosing an on-shell scheme, with $\mu_r^2=-m^2$. The latter being more physical (in the sense that it parametrizes the model in terms of a renormalization group invariant quantity, the pole of the propagator), this suggests that the polynomial approach to the quenched limit in a generic scheme may be a mere approximation artifact. In this work, we shall not implement the on-shell scheme for the gluon because we would like to stick to the scheme used in  Ref.~\cite{reinosa2016two} for simplicity. Moreover, it is not completely clear to us what needs to be done if the pole becomes complex \cite{Gribov77,Dudal:2010wn,Siringo:2016jrc}. 

Similar remarks apply to the fermion contribution in the last two lines of \Eqn{eq:fgegf}. We could implement an on-shell scheme for the quark sector by choosing $\hat\mu_f^2=\mu^2_{f}=-M^2_f$ as a simple way to account for higher order contributions. But, again, it is not clear what to do when the pole becomes complex. In what follows, we show results in the scheme $\hat\mu_{f}=\mu_r$ and $\mu_f=0$ and briefly discuss how our results vary when changing the values of $\hat\mu_f$ and $\mu_f$.

\subsection{Symmetries}
The background field potential in the LDW gauge possesses various symmetries, discussed in detail in Refs.~\cite{reinosa2015perturbative,reinosa2016two}, which provide a useful cross-check of our two-loop calculation. The two-loop pure glue and one-loop quark contributions have already been discussed in these references so we shall concentrate here on the two-loop quark sunset contribution. We shall use that the background $r$ only enters in shifts of Matsubara frequencies through scalar products either with roots, $\pm iT\,r\cdot\alpha$, for bosonic ones or with weights, $\pm iT\,r\cdot\rho$, for fermionic ones. Clearly, shifting bosonic or fermionic Matsubara frequencies by a multiple of $2\pi T$ can be reabsorbed by a change of summation variables and does not affect the result.\footnote{A similar observation can be made on the expressions where Matsubara sums have been explicitly performed. In that case, the background $r$ enters the final expression either in bosonic thermal factors, as $n_{\varepsilon_{m,q}\pm iT r\cdot\alpha}$ and $n_{q\pm iT r\cdot\alpha}$, or in fermionic ones, as $f_{\varepsilon_{M,q}\pm(\mu-iTr\cdot\rho)}$, where $\varepsilon_{m,q}\equiv\sqrt{q^2+m^2}$ and $n_x=[\exp(\beta x)-1]^{-1}$ and $f_x=[\exp(\beta x)+1]^{-1}$ are the Bose-Einstein and Fermi-Dirac distribution functions, respectively. These are invariant under an imaginary shift of $2\pi T$: $n_{x+2i\pi T}=n_x$, and similarly for $f_x$.} 

First of all, the potential is invariant under color rotations that leave the background of the form \eqn{eq:bg0}. These are the so-called Weyl transformations, which correspond to reflections with respect to axes orthogonal to the roots \cite{reinosa2016two,Zuber}. As can be seen on Fig.~\ref{fig:root}, these result in permutations of roots and of weights. Now, the two-loop contribution \eqn{eq:res} to the potential involves either simple sums over roots $\alpha$ or weights $\rho$, or sums over color conserving triplets $(\alpha,\beta,\gamma)$ or $(\sigma,\rho,\alpha)$ such that $\alpha+\beta+\gamma=0$ and $\sigma=\rho+\alpha$, respectively. The simple sums are trivially left unchanged under the Weyl transformation and it is simple to check that color conserving triplets are permuted to one another so that the sums over such triplets are also left invariant.

Next, the potential is invariant under periodic gauge transformations that preserve the form of the background (\ref{eq:bg0}). These correspond to translations of the vector $r$ by $4\pi\alpha'$ where $\alpha'$ is any root \cite{reinosa2016two}, which shift the scalar products $r\cdot\alpha$ and $r\cdot\rho$ by $4\pi\alpha'\cdot\alpha$ and $4\pi\alpha'\cdot\rho$, respectively. It is easily checked that these are multiples of $2\pi$ for all possible roots and weights. It follows from the remarks made above that this leaves the expression \eqn{eq:res} invariant.

The pure YM part of the potential is also invariant under twisted gauge transformations that preserve the form of the background \eqn{eq:bg0}. They correspond to translations of the background by $4\pi\rho'$ where $\rho'$ is any weight and since the scalar products $r\cdot\alpha$---the only ones entering the pure YM potential---are shifted by multiples of $2\pi$, see above, this part of the potential is invariant. On the other hand, the scalar products $r\cdot\rho$ that enter the quark contribution are shifted by $4\pi\rho'\cdot\rho$. These shifts are either $4\pi/3$ or $-2\pi/3$ and, thus, do not leave the quark contribution to the potential invariant. This is the manifestation of the explicit breaking of center symmetry by fundamental quarks. However, making a simultaneous shift of the chemical potential by $2i\pi/3$ results in a total shift of the Matsubara frequencies by a multiple of $2\pi$. It follows that the full two-loop potential is invariant under center transformations provided one simultaneously changes $\mu$ to $\mu+2i\pi/3$. This is the so-called Roberge-Weiss symmetry.

Another important symmetry is charge conjugation. It implies that the potential is invariant under the simultaneous transformation $r\to -r$ and $\mu\to -\mu$. This can be easily checked by using the identities
\begin{align}
J_m^\kappa(-r)&=J_m^{-\kappa}(r)=J_m^\kappa(r)\,, \\
\tilde J_m^\kappa(-r)&=\tilde J_m^{-\kappa}(r)=-\tilde J_m^\kappa(r)\,, 
\end{align}
\begin{align}
J_M^\rho(-r,\mu)&= J_M^{-\rho}(r,\mu)=J_M^\rho(r,-\mu)\,,  \\
\tilde J_M^\rho(-r,\mu)&=J_M^{-\rho}(r,\mu)=-\tilde J_M^\rho(r,-\mu)\,,
\end{align}
and
\beq
S_{mMM}^{\kappa\sigma\rho}(r,\mu)=S_{mMM}^{(-\kappa)(-\sigma)(-\rho)}(-r,\mu)=S_{mMM}^{\kappa\sigma\rho}(r,-\mu)\,.
\eeq

Finally, the potential is invariant under complex conjugation provided one changes simultaneously $r\to r^*$ and $\mu\to-\mu^*$. Again, this is easily checked using
\begin{align}
J_m^\kappa(r)^*&=J_m^\kappa(r^*)\,,\\
 \tilde J_m^\kappa(r)^*&=\tilde J_m^\kappa(r^*)\,,
\end{align}
\begin{align}
J_M^\rho(r,\mu)^*&=J_M^\rho(r^*,-\mu^*)\,,\\
\tilde J_M^\rho(r,\mu)^*&=\tilde J_M^\rho(r^*,-\mu^*)\,,
\end{align}
and
\beq
S_{mMM}^{\kappa\rho\sigma}(r,\mu)^*=S_{mMM}^{\kappa\rho\sigma}(r^*,-\mu^*)\,.
\eeq
In particular, we verify that, for imaginary chemical potential, the potential is real for $r=(r_3,r_8)\in\mathds{R}\times\mathds{R}$, whereas, for real chemical potential, it is real for $r=(r_3,r_8)\in\mathds{R}\times i\mathds{R}$ (this last property requires, in addition to complex conjugation, charge conjugation and the Weyl symmetry $r_3\to -r_3$).

\section{Results}

\label{sec:results}

We are now ready to compute the two-loop corrections to the results of Ref.~\cite{reinosa2015perturbative} concerning the phase diagram in the $(T,\mu)$-plane. We compare our results to lattice data and to those of other, nonperturbative continuum approaches. We study separately the cases of vanishing, imaginary, and real chemical potential. In the latter case, we discuss and interpret the behavior of the Polyakov loops \eqn{eq_popoldef} and \eqn{eq_popoldefbar} as functions of $\mu$. We also investigate various thermodynamical observables, such as the pressure, the energy density, or the entropy, and we study the thermodynamical stability at the present order of approximation. Two-loop contributions have been shown before to cure certain spurious unphysical features of the leading-order thermodynamics in the pure glue case \cite{Reinosa:2014zta,reinosa2016two}. 

Int what follows, we set the scale $\mu_r=1$~GeV. For simplicity, as we vary the temperature, chemical potential, and quark masses, we keep fixed the values $g=4.9$ and $m=540$~MeV, determined from fits of the one-loop vacuum propagators of the pure YM theory against lattice data \cite{Tissier:2011ey}. This eases the comparison with Ref.~\cite{reinosa2016two}, where the same set of parameters was used.

\subsection{Vanishing chemical potential}
\label{ZeroMu}
At zero chemical potential, charge conjugation symmetry is manifest and we can restrict the background field potential to the median $r_8=0$ of the fundamental Weyl chamber, which is the locus of charge invariant backgrounds in this Weyl chamber \cite{reinosa2016two}. This means that it is enough to take backgrounds of the form $\smash{r=(r_3,0)}$ with $r_3\in [0,2\pi]$. As already mentioned in Sec.~\ref{sec:add_quarks}, the physical value of the background is obtained as the absolute minimum of the potential. 

\begin{figure}
	\centering
	\includegraphics[width=0.45\textwidth]{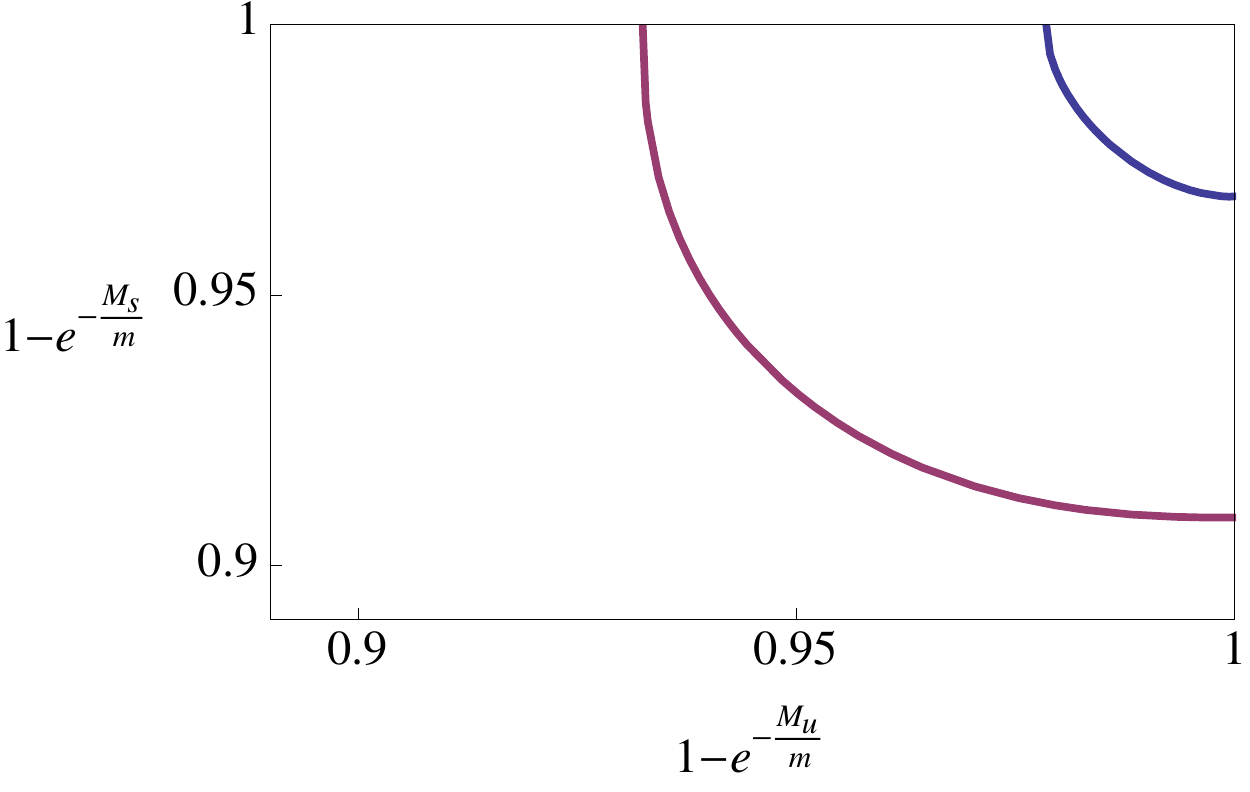}
	\caption{\label{fig:Columbia}Top right corner of the Columbia plot at one-loop (purple) and two-loop (blue) orders of our approach.}
\end{figure}

In the limit of infinite quark masses (pure YM), the minimum of the potential plays the role of an order parameter for center symmetry, with a first order phase transition as a function of the temperature. In the presence of large but finite quark masses, center symmetry is explicitly broken. However, the minimum of the potential still experiences a first order jump, at least as long as the quark masses are large enough. As these masses are decreased, this first order transition turns into a crossover. In Fig.~\ref{fig:Columbia}, we show the boundary line between these two regimes for $2+1$ flavors in the plane $\smash{(M_u=M_d,M_s)}$. We observe that, for the considered renormalization scheme (corresponding to $\smash{\hat\mu_f=\mu_r}$ and $\smash{\mu_f=0}$, see above), the boundary line is pushed towards larger masses (the region of first order phase transitions shrinks) as compared to the one-loop case. The same behavior is observed using other renormalization schemes. The critical temperature along the boundary line is found to be almost independent\footnote{We find $\frac{T_c(N_f=3)-T_c(N_f=1) }{T_c(N_f=1)}\approx 0.2 \%$. This is significantly smaller than the difference with the pure YM case: $\frac{T_c(N_f=1)-T_c(N_f=0) }{T_c(N_f=0)}\approx 3.8 \%$. The latter is due to the fact that the typical fermion mass on the critical line is $M_c/T_c\sim 7-8$ which leads to a significative Boltzmann suppression} of $N_f$ and equals  $T_c/m\approx0.456$, a slightly smaller value than the one obtained in the pure YM case at the same order of approximation, $T_c/m\approx0.474$ \cite{reinosa2016two}. So, as it was already the case at one-loop order, introducing quarks reduces the critical temperature.

In the one-loop calculation of Ref.~\cite{reinosa2015perturbative}, the ratio
\beq\label{eq:ratioR}
R_{N_f}\equiv\frac{M_c(N_f)}{T_c(N_f)}\,,
\eeq
was evaluated for three points along the critical line in the Columbia plot, corresponding to $N_f=1$, $2$ and $3$ degenerate flavors, and compared to similar ratios computed on the lattice, with a very good agreement, given the simplicity of the calculation. The direct comparison to lattice results made sense because, at one-loop order, the renormalized mass parameter $M$ coincides with the bare mass and is therefore scheme independent to this order. In fact, lattice results are expressed in terms of the bare mass as well \cite{Fromm:2011qi}. Beyond leading order, such a comparison becomes difficult, if not impossible, because the ratios \eqn{eq:ratioR} become scheme dependent \cite{Fischer:2014vxa}. However, a meaningful comparison to lattice results is still possible  thanks to the following observation. At the present order of approximation, the mass renormalization factor, $Z_M=M_{\rm b}/M$, which relates the bare and renormalized masses, is obtained from the one-loop contribution to the quark self-energy. The latter is trivially independent of $N_f$ because no fermion loop is involved. It follows that ratios of the form $M_c(N_f')/M_c(N_f)$ are scheme independent up to higher order corrections. Given the fact that $T_c$ is essentially independent of $N_f$, it thus makes sense to compare the two-loop values of the ratios $R_{N_f'}/R_{N_f}\approx M_c(N_f')/M_c(N_f)$ with the lattice results.

In Table \ref{ZeroCritics}, we show our results for $R_{N_f}$ at one- and two-loop orders using the renormalization scheme $\hat\mu_f=\mu_r$ and $\mu_f=0$, and we compare them to the lattice results of Ref.~\cite{Fromm:2011qi}. We also quote the results of other continuum approaches, namely, the DSE calculation of Ref.~\cite{Fischer:2014vxa} and the matrix model of Ref.~\cite{Kashiwa:2012wa}. As pointed out in Ref.~\cite{reinosa2015perturbative}, the one-loop results are in very good agreement with the lattice ones and we observe that, if the two-loop corrections do not significantly alter this conclusion, they seem to worsen the agreement in some cases. As explained above, due to the scheme dependence of the quantities $R_{N_f}$, such a direct comparison must not be taken too seriously (neither must the apparently bad agreement of DSE results with the lattice ones \cite{Fischer:2014vxa}). If, instead, we compare the ratios $\smash{M_c(N_f)/M_c(N_f=1)}\approx R_{N_f}/R_1$, we observe, first, a pretty good agreement with the lattice values and, second, the fact that the two-loop corrections improve the one-loop results.

Although our argument for considering the ratios $R_{N_f}/R_1$ relies on perturbation theory, it is interesting to compute these ratios from the results of the other continuum approaches, which is also presented in Table~\ref{ZeroCritics}. The effect of partially eliminating the scheme dependence is most remarkable for the DSE results, although the agreement is not as good as for our perturbative results. This may be due to the fact that the DSE results are not exactly two-loop: they contain only part of the two-loop corrections and include partially some higher order corrections, in which case the scheme dependence may be more important. Finally, we note the excellent agreement of the matrix model with the lattice results.

\begin{table}[h!]
  \centering
	\begin{tabular}{|c | c c c ||| c  c|} 
		\hline
		$R_{N_f}$ & $N_f=1$ & $N_f=2$  &$N_f=3$ & $R_2/R_1$ & $R_3/R_1$ \\ [0.5ex] 
		\hline\hline
		1-loop & 6.74 & 7.59 & 8.07 &1.12 & 1.20 \\ 
		\hline
		2-loop & 7.53 & 8.40 & 8.90 & 1.11 & 1.18 \\
		\hline
		Lattice & 7.23 & 7.92 &  8.33 & 1.10 & 1.15 \\
		\hline
		DSE \cite{Fischer:2014vxa} & 1.42 & 1.83 & 2.04 & 1.29 & 1.43\\
		\hline
	Matrix \cite{Kashiwa:2012wa} & 8.04 & 8.85 & 9.33 & 1.10 & 1.16\\
		\hline
	\end{tabular}
	\caption{Left side of the table: our results for the ratio $M_c/T_c$ on the critical line of the Columbia plot for $\smash{N_f=1}$, $2$, and $3$ degenerate flavors, computed in the scheme $\smash{\hat\mu_f=\mu_r}$ and $\smash{\mu_f=0}$, and compared to other approaches. Right side of the table: comparison of the ratios $R_{N_f}/R_1$, less sensitive to scheme dependences, as explained in the main text.}
	\label{ZeroCritics}
\end{table}

We have studied other schemes by varying $\hat\mu_f$ and $\mu_f$ (while keeping however $m$ and $g$ fixed). We observe that scheme dependences are not as small as expected, specially as we increase $\mu_f$ (the effect is much smaller for $\hat\mu_f$). We understand this from the fact that, as the renormalization scale is varied over a wide range, it is not justified to neglect renormalization group effects.

\subsection{Imaginary chemical potential}

The case of an imaginary chemical potential $\mu\equiv i\mu_i\equiv iT\tilde\mu_i$ is of interest because it can easily be simulated with lattice techniques since the sign problem is absent. In the present approach, this also guarantees that the physical value of the background corresponds to the absolute minimum of the potential \cite{Reinosa:2016xaj}. However, the charge conjugation symmetry being explicitly broken by the chemical potential, the latter typically moves away from the $r_8=0$ axis in the Weyl chamber.

We obtain the same qualitative features as in the one-loop case, also in agreement with lattice results. For large temperatures, the system displays a first order phase transition (a.k.a the Roberge-Weiss transition) when $\mu_i$ crosses $\pi/3$. In the $(\mu_i,T)$ plane and for sufficiently large quark masses, the Roberge-Weiss transition line at $\tilde\mu_i=\pi/3$ is connected by a line of first order transition to the corresponding transition point at $\tilde\mu_i=0$. The diagram being symmetric about $\tilde\mu_i=\pi/3$, a similar curve connects the Roberge-Weiss transition to a first-order phase transition point at $\tilde\mu_i=2\pi/3$. As the quark mass is decreased, the first-order phase transitions at $\tilde\mu_i=0$ and $\tilde\mu_i=2\pi/3$ become second order ones when $M$ reaches the critical value $M_c(\mu_i=0)$. As the quark mass is further decreased, the corresponding critical points move inside the $(T,\mu_i)$-plane towards $\tilde\mu_i>0$ and $\tilde\mu_i<2\pi/3$,  respectively. These two points eventually merge into a tricritical point at $\tilde\mu_i=\pi/3$ when $M$ reaches a (tri)critical value $M_c(\mu_i=\pi T/3)$.

For a more quantitative comparison, we again use the ratios
\begin{equation}
\frac{M_c(N_f,\tilde\mu_i)}{M_c(N_f=1,\tilde\mu_i)}\approx \frac{R_{N_f}(\tilde\mu_i)}{R_1(\tilde\mu_i)}\,,
\end{equation}
where we have made the dependence on the (imaginary part of the) chemical potential explicit, as well as
\begin{equation} \label{secondRatios}
\frac{M_c(N_f,0)}{M_c(N_f,\pi/3)}\approx \frac{R_{N_f}(0)}{R_{N_f}(\pi/3)}.
\end{equation}
The justifications for using the latter are, first, that the mass renormalization factor $Z_M$ introduced in the previous section is not only independent of $N_f$, but also of $T$ and $\mu$ (in the vacuum renormalization scheme considered here), and, second, that $T_c(\mu_i)$ does not vary much over the range $\tilde\mu_i\in [0,\pi/3]$ (although we note that the relative variation $\approx 1\%$ is not as small as the one observed when varying $N_f$ at $\smash{\mu=0}$). Our results are gathered in Tables \ref{ImagCritics} and \ref{ImagCritics2} and compared to those of other approaches. We note that our results for the ratios $R_{N_f}(0)/R_{N_f}(\pi/3)$ are less conclusive since the relative variations observed between the one- and two-loop results are of the same order than the relative variations of $T_c$ as $\tilde\mu_i$ is varied between $0$ and $\pi/3$. To eliminate this effect, one should directly compare mass ratios $M_c(N_f,0)/M_c(N_f,\pi/3)$. Unfortunately, these are not available in the lattice literature.

\begin{table}[h!]

\begin{center}
\begin{tabular}{|c | c c c ||| c c|}
\hline
$R_{N_f}(\pi/3)$ & $N_f=1$ & $N_f=2$ &$N_f=3$ & $R_2/R_1$ & $R_3/R_1$ \\ [0.5ex]
\hline\hline
1-loop \cite{reinosa2015perturbative} & 4.74 & 5.63 & 6.15 & 1.19 & 1.30 \\
\hline
2-loop & 5.47 & 6.41 & 6.94 & 1.17 & 1.27 \\
\hline
Lattice \cite{Fromm:2011qi} & 5.56 & 6.25 & 6.66 & 1.12 & 1.20 \\
\hline
DSE \cite{Fischer:2014vxa} & 0.41 & 0.85 & 1.11 & 2.07 & 2.70 \\
\hline
Matrix \cite{Kashiwa:2012wa} & 5.00 & 5.90 & 6.40 & 1.18 & 1.28 \\
\hline
\end{tabular}
\caption{Left side of the table: our results for the ratio $M_c/T_c$ on the boundary line of the Columbia plot at $\smash{\tilde\mu_i=\pi/3}$ for $N_f=1$, $2$ and $3$ degenerate flavors, computed in the scheme $\smash{\hat\mu_f=\mu_r}$ and $\smash{\mu_f=0}$, and compared to lattice results and to other continuum approaches. Right side of the table: comparison of the ratios $R_{N_f}(\pi/3)/R_1(\pi/3)$, less sensitive to scheme dependences, as explained in the main text.}
\label{ImagCritics}
\end{center}

\end{table}

\begin{table}[h!]

\begin{center}
\begin{tabular}{|c | c c c |}
\hline
$R_{N_f}(0)/R_{N_f}(\pi/3)$ & $N_f=1$ & $N_f=2$ &$N_f=3$ \\ [0.5ex]
\hline\hline
1-loop \cite{reinosa2015perturbative} & 1.42 & 1.35 & 1.31 \\
\hline
2-loop & 1.38 & 1.31 & 1.28 \\
\hline
Lattice \cite{Fromm:2011qi} & 1.30 & 1.27 & 1.25 \\
\hline
DSE \cite{Fischer:2014vxa} & 3.46 & 2.15 & 1.84 \\
\hline
Matrix \cite{Kashiwa:2012wa} & 1.61 & 1.50 & 1.46 \\
\hline
\end{tabular}
\caption{The ratios $R_{N_f}(0)/R_{N_f}(\pi/3)$ at one- and two-loop orders, compared to lattice results and to other continuum approaches.}
\label{ImagCritics2}
\end{center}

\end{table}

As pointed out in Ref.~\cite{deForcrand:2010he}, the vicinity of the tricritical point is approximately described by the mean field scaling behavior
\begin{equation}
\frac{M_c(\mu_i)}{T_c(\mu_i)}=\frac{M_{\rm tric.}}{T_{\rm tric.}}+K\left[ \left(\frac{\pi}{3}\right)^2 -\left(\frac{\mu_i}{T_c}\right)^2 \right]^{\frac{2}{5}}\,.
\end{equation} 
\begin{figure}
	\centering
	\includegraphics[width=0.45\textwidth]{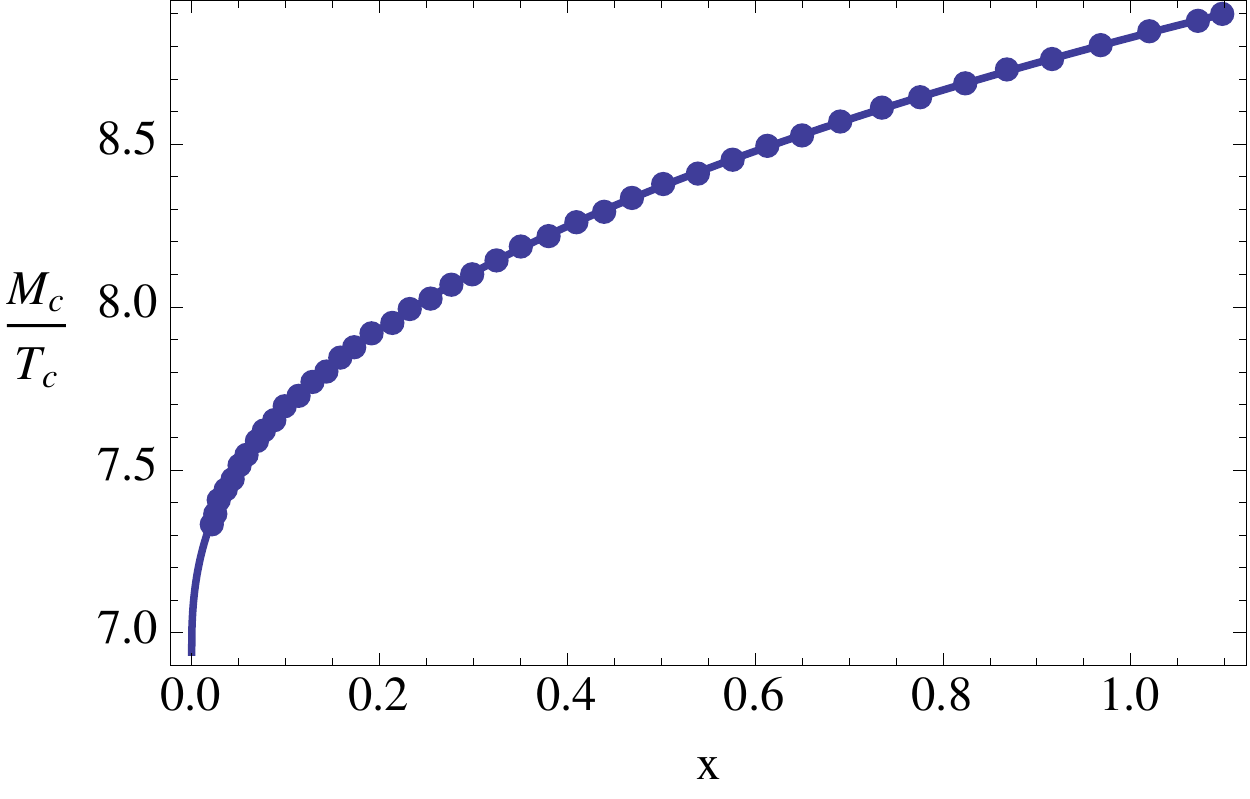}
	\includegraphics[width=0.45\textwidth]{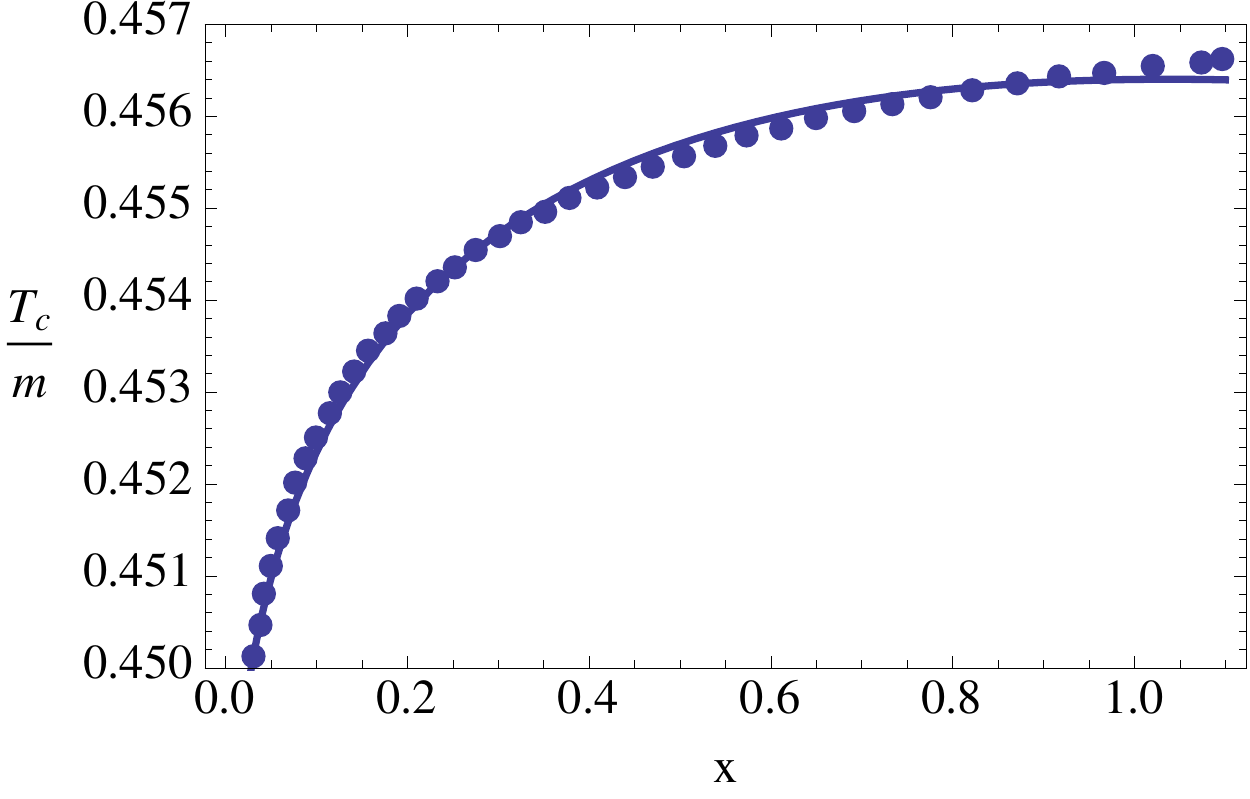}
	\caption{\label{fig:ImagScaling} Top panel: Quark mass to temperature ratio at the critical point as a function of  $x\equiv(\pi/3)^2+(\mu/T_c)^2=(\pi/3)^2-(\mu_i/T_c)^2$ for $N_f=3$ degenerate flavors.
		Bottom panel: The temperature at the critical point in units of the gluon mass parameter $m$ as a function of $x$ and for $N_f=3$.}
\end{figure}
We track the $Z_2$ critical points in the plane $(T,\mu_i)$ for (degenerate) quark mass $M\in[M_c(\pi T/3),M_c(0)]$ for $N_f=3$. Our results, shown in Fig.~\ref{fig:ImagScaling}, are well described by the following fits to the tricritical scaling expressions:
\begin{eqnarray}
\frac{M_c}{T_c}(x)&\approx& 6.939 + 1.888\, x^{2/5}\,,\label{tricScaling}\\
\frac{T_c}{m}(x)&\approx& 0.445 + 0.022\, x^{2/5} - 0.011\, x^{4/5}\,,
\end{eqnarray} 
with $m$ is the gluon mass parameter and $x\equiv(\pi/3)^2-(\mu_i/T_c)^2$.

\subsection{Real chemical potential}

As pointed out in Ref.~\cite{reinosa2015perturbative}---see also Sec.~\ref{sec:add_quarks}---, for a real chemical potential, the component $r_8$ of the background needs to be continued from $\mathds{R}$ to $i\mathds{R}$. This ensures that the background field potential remains real (it becomes complex if one insists on keeping $r_8\in\mathds{R}$) and, in turn, that thermodynamical quantities extracted from the latter are real as well. An inconvenience of this continuation is however that the physical value of the background $(r_3,r_8)$ is not associated with the absolute minimum of the potential anymore, but rather with a saddle point in the space $\mathds{R}\times i\mathds{R}$; and that there exists no clear rule (yet) as to which saddle actually corresponds to the physical point if there are more than one present. In what follows, we follow the recipe of Ref.~\cite{reinosa2015perturbative} and we always select the deepest saddle point.\footnote{This rule is valid at $\mu=0$ where one can work equivalently over $\mathds{R}\times\mathds{R}$ or $\mathds{R}\times i\mathds{R}$ and the physical point lies at $r_8=0$. The absolute minimum in the first subspace should coincide in this case with the deepest saddle point in the second subspace.}

\begin{figure}
	\centering
	\includegraphics[width=0.45\textwidth]{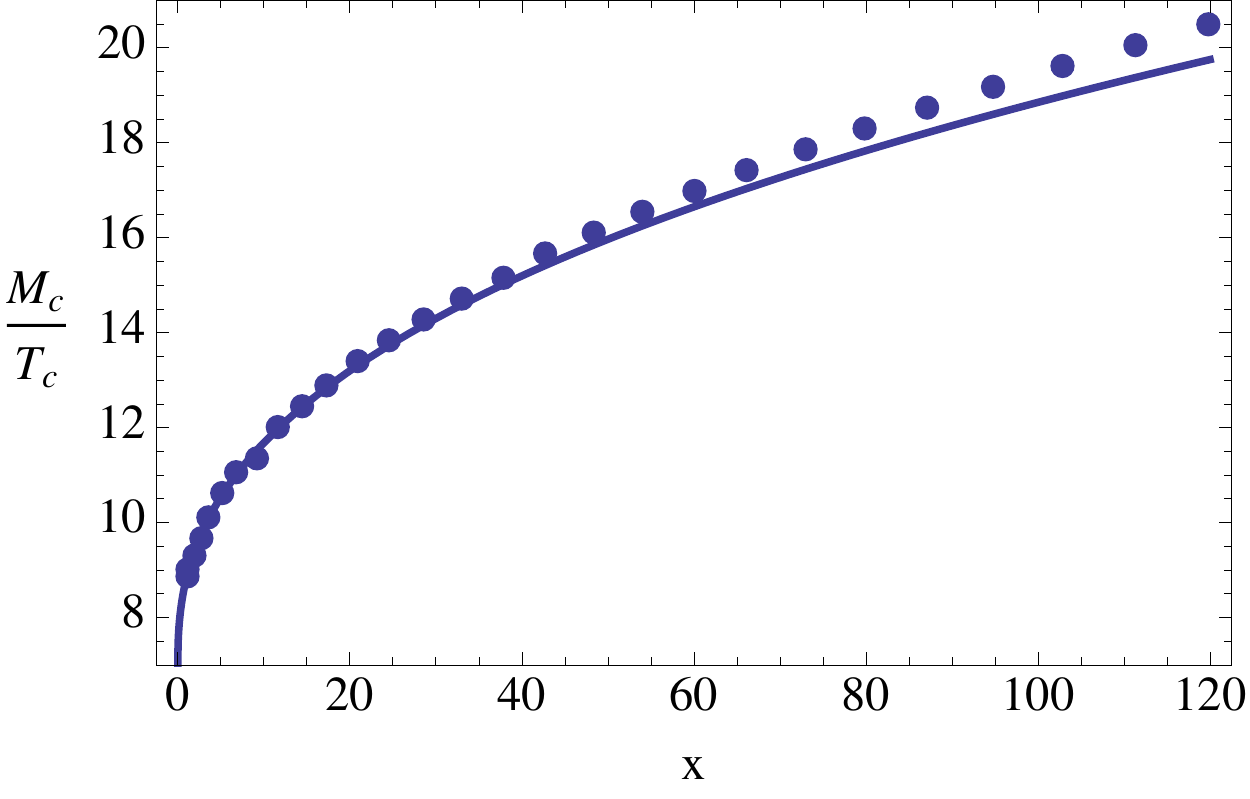}
	\caption{\label{fig:RealScaling} Tricritical scaling extrapolated to the region $\mu^2>0$.}
\end{figure}

In line with the one-loop results of Ref.~\cite{reinosa2015perturbative}, we find that the top-right critical boundary in the Columbia plot shrinks towards the YM point for increasing values of the chemical potential.  We also plot, in Fig \ref{fig:RealScaling}, the $\mu$ dependence of the ratio $M_c/T_c$ on the critical boundary and compare it to the tricritical scaling \eqn{tricScaling}, continued to $\mu\in\mathds{R}$.
We observe that, as was the case at one-loop, this describes the data well up to rather large values of $x=(\pi/3)^2+(\mu/T_c)^2 \lesssim 40$. A $\chi^2$-analysis shows that the extrapolation of the fit to real $\mu$ describes the two-loop results better than the one-loop one, although the one-loop extrapolation was already quite good.\\

 \begin{figure}[t]
	\centering
	\includegraphics[width=0.46\textwidth]{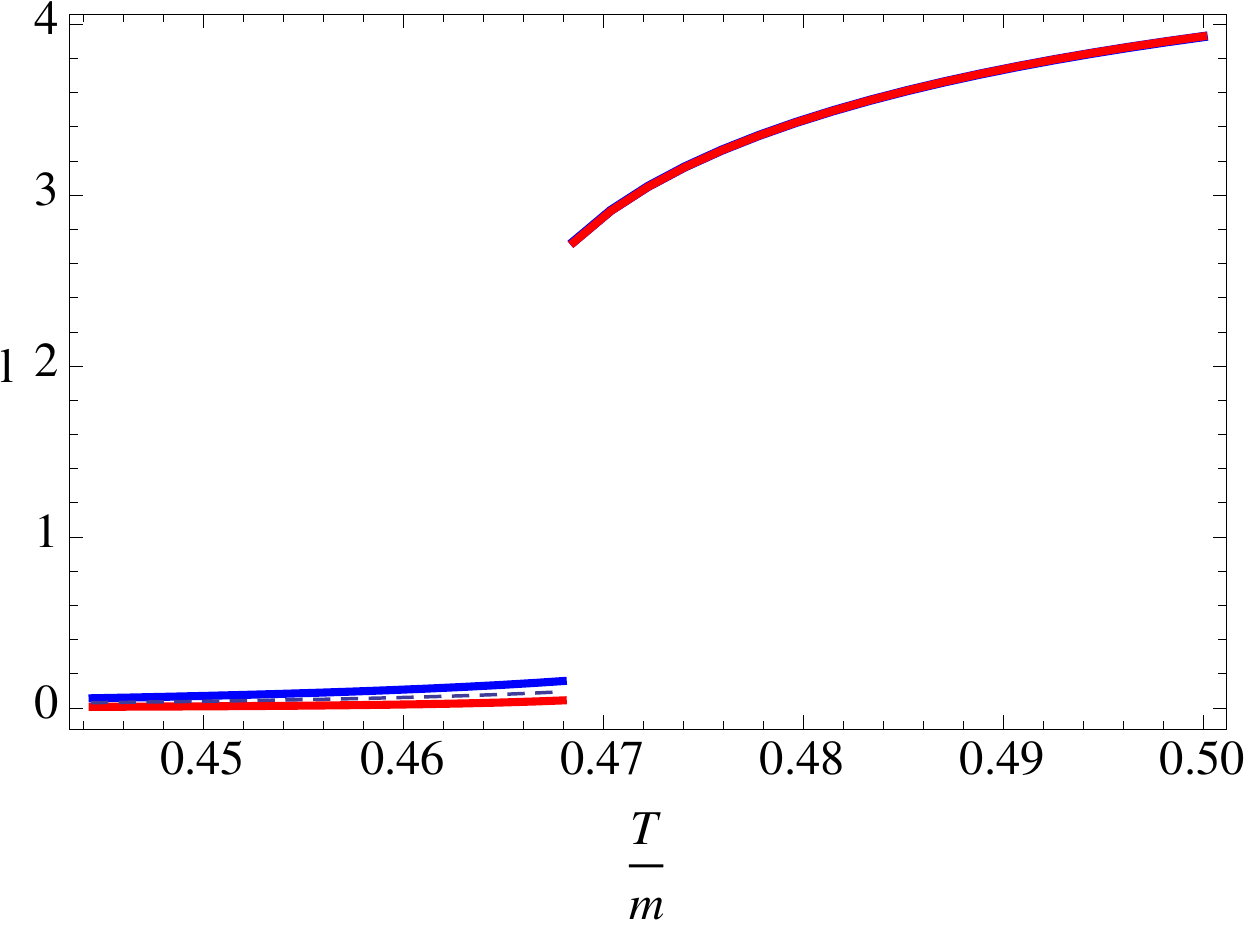}\\
	\vglue5mm
	\includegraphics[width=0.46\textwidth]{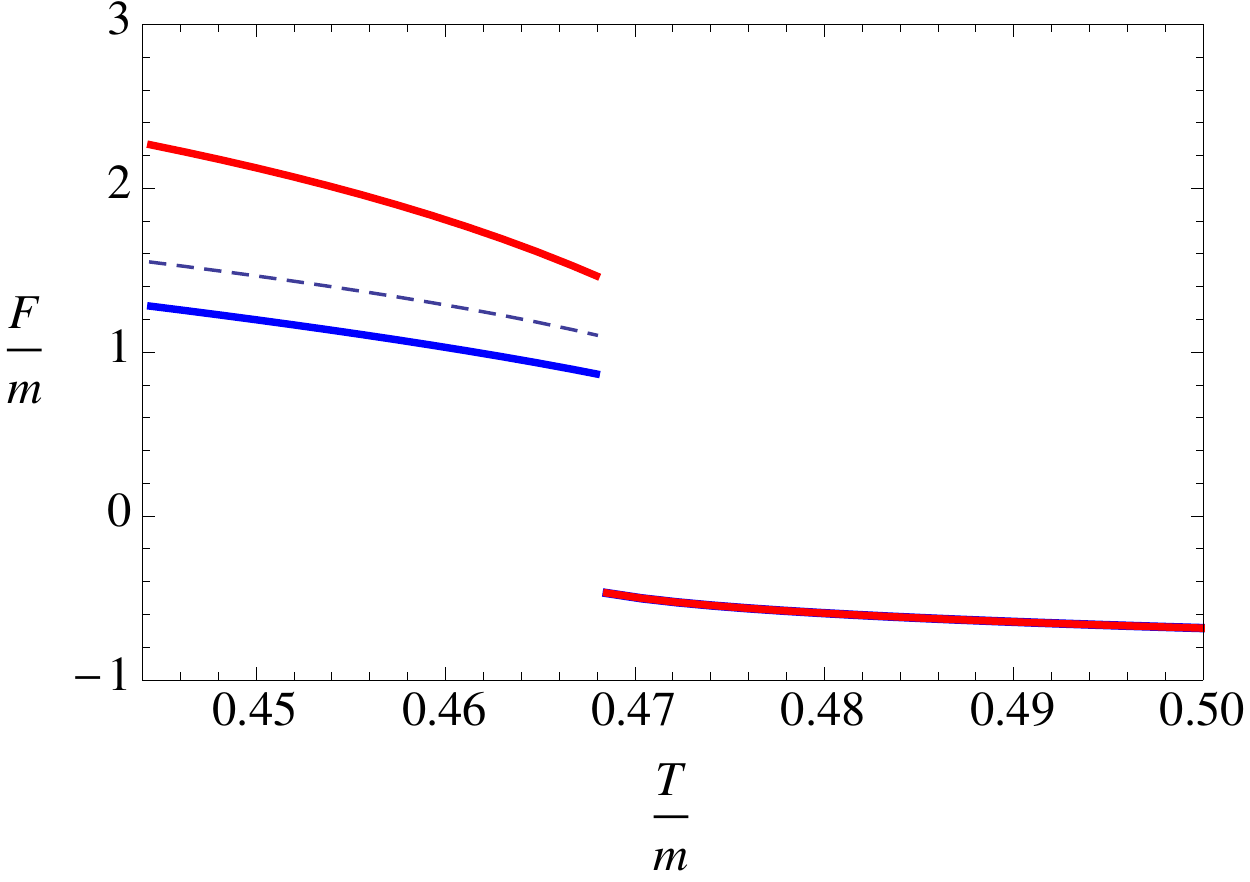}
	\caption{\label{fig:PressureEntropy} The Polyakov loops $\ell$ (blue) and $\bar\ell$ (red) and the corresponding quark and anti-quark free energies computed with a nonzero imaginary background along the $r_8$ direction. We also show the same quantities computed with $r_8 =0$ (and $r_3$ at the minimum of the  potential along this axis). }
\end{figure}

In Fig.~\ref{fig:PressureEntropy}, we show the Polyakov loops $\ell$ and $\bar\ell$ as functions of the temperature, together with the corresponding (static) quark and anti-quark free energies for $N_f=3$ degenerate flavors with $\mu/m=0.6$ and $M/m=5.56$, in which case the transition is of first order. We compare these quantities to those obtained by evaluating the Polyakov loops at the minimum of the potential along the axis $r_8=0$, as done, e.g., in Ref.~\cite{Fischer:2014vxa}. We insist that this does not correspond to an extremum of the potential and is, thus, at best, an approximation for not too large $\mu$. Indeed, we see that there can be substantial differences with the loops evaluated at the saddle points for what concerns the static (anti)quark free energies in the confined phase. In particular, taking $\smash{r_8=0}$ completely misses the fact that the chemical potential introduces a difference between quarks and anti-quarks. We mention that this difference is more visible in the confined phase and that, for $\mu>0$, corresponding in our convention to an excess of anti-quarks with respect to quarks, we find that it costs less energy to bring a quark than an anti-quark to the bath. This in line with observations made on the lattice and can be interpreted in terms of screening of a quark by the anti-quarks of the thermal bath \cite{Kaczmarek:2002mc}.
 \begin{figure}[t]
	\centering
	\includegraphics[width=0.46\textwidth]{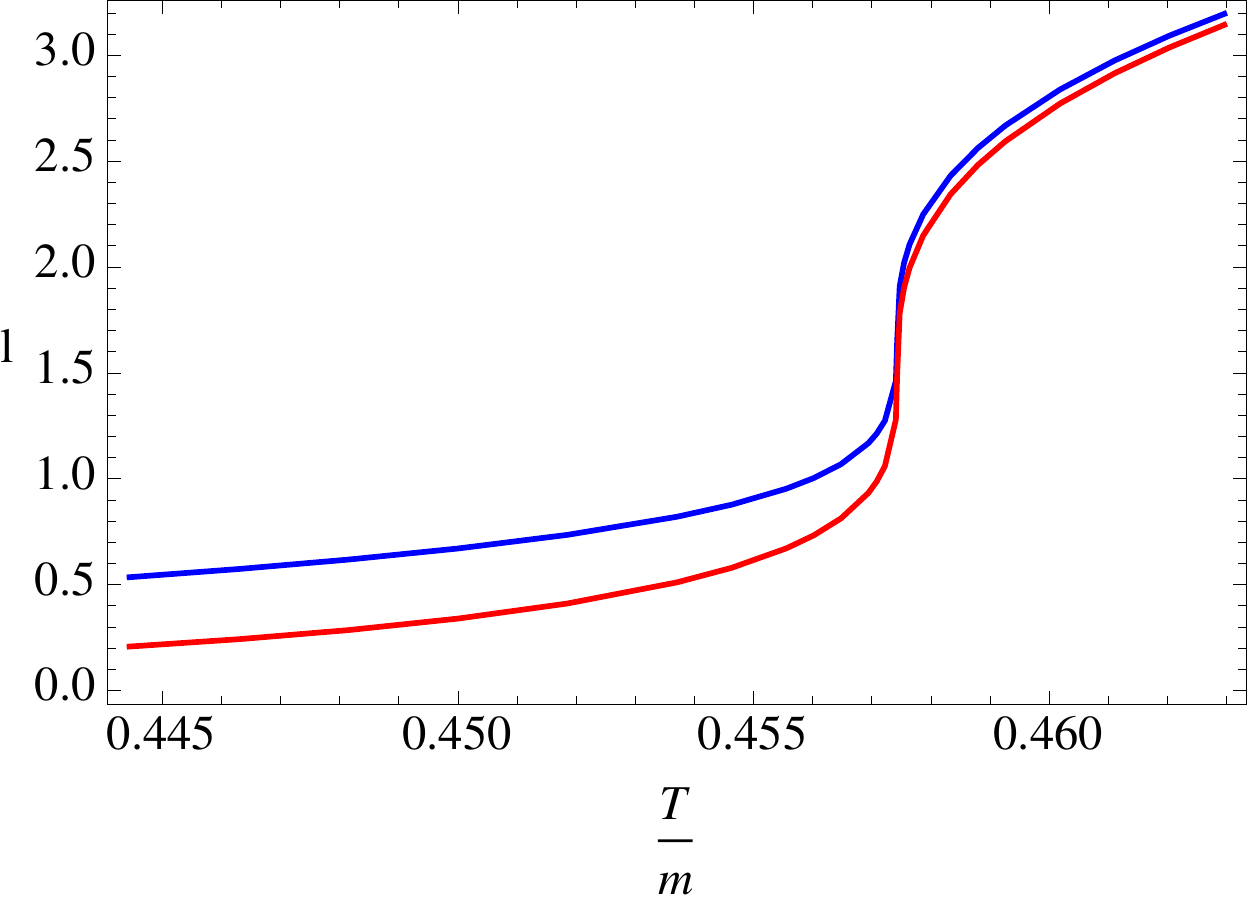}\\
	\vglue5mm
	\includegraphics[width=0.46\textwidth]{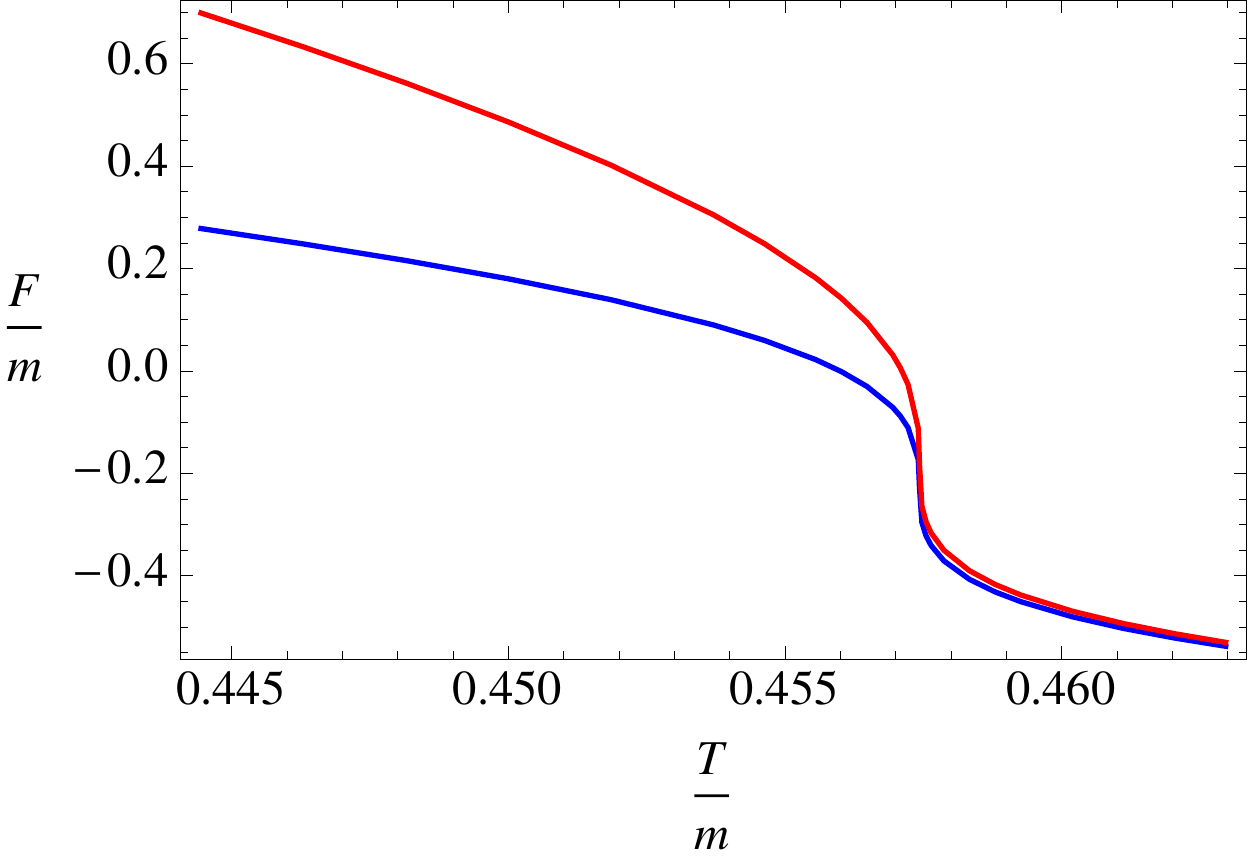}
	\caption{\label{fig:PressureEntropy-critical} Same as in \Fig{fig:PressureEntropy} for parameters corresponding to the critical line in the Columbia plot. }
\end{figure}
For completeness, we also show the Polyakov loops and the corresponding free energies as functions of $T$ at the critical point $M=M_c$ for $3$ degenerate flavors and    $\mu/m=0.6$. The effect of the nonzero chemical potential mentioned above is more pronounced in both phases.

We also remark from Figs.~\ref{fig:PressureEntropy} and \ref{fig:PressureEntropy-critical} that the Polyakov loops overshoot $1$ in the high temperature phase, as was also observed for the pure YM theory \cite{Reinosa:2016xaj}. In the later case, this clearly signals an artifact of the perturbative expansion. Indeed the Polyakov loop $\ell$ is the average of ${\rm tr}\,L/N$ where $L$ is the path ordered exponential along the compact time direction.\footnote{In this work, we are considering the bare Polyakov loop since, at the present order of approximation, in dimensional regularization, it is UV finite. On the contrary renormalized Polyakov loops \cite{Kaczmarek:2002mc} have no reason to be bounded by $1$.} The latter being unitary, we have $|{\rm tr}\,L/N|\le1$. Moreover, the integration measure under the functional integral, including the fermion determinant, being positive, it follows that $|\ell|<1$. So, the result $\ell>1$ in the pure YM case could originate either from the fact that the perturbative expansion violates the unitarity of $L$ or that the necessary gauge-fixing usually modifies the integration measure into a nonpositive definite one.\footnote{A notorious exception is the Gribov-Zwanziger implementation of the Langau gauge.}

The previous discussion extends trivially in the presence of quarks with an imaginary chemical potential. In contrast, for real chemical potential, the discussion becomes more intricate, in part due to the fact that the fermion determinant becomes complex (signalling once again that the sign problem persists in some form in continuum approaches) and there is no reason a priori for $\ell$ to be bounded by $1$ anymore. In particular, it is difficult to decide what part of $\ell$ is due to an artifcat of the perturbative expansion. Still, observing bare Polyakov loops above one, even in the case of a real chemical potential, seems in contradiction with the interpretation of their logarithms as minus $\beta$ times the free energy differences for having a quark or an anti-quark with respect to the vacuum \cite{Svetitsky:1985ye}. Indeed, finding a bare Polyakov loop above one, would mean that these free energy differences are negative. It has been argued, however, see Ref.~\cite{Kaczmarek:2002mc}, that one should instead define renormalized Polyakov loops\footnote{The bare Polyakov loops have a vanishing limit on the lattice due to divergences in the corresponding logarithms.} for which the previous interpretation does not apply.

\begin{figure}[t]
	\centering
	\includegraphics[width=0.46\textwidth]{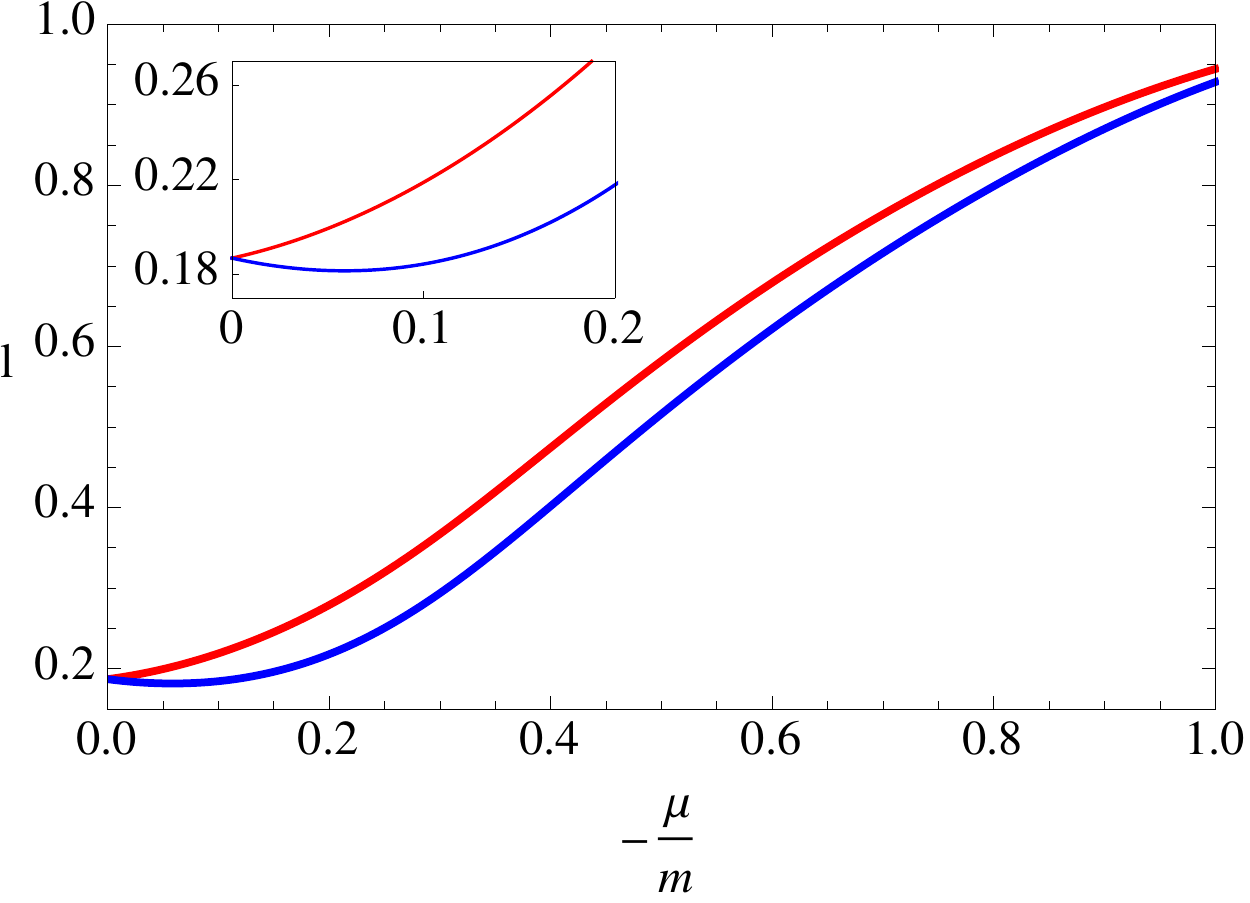}\\
	\vglue5mm
	\includegraphics[width=0.46\textwidth]{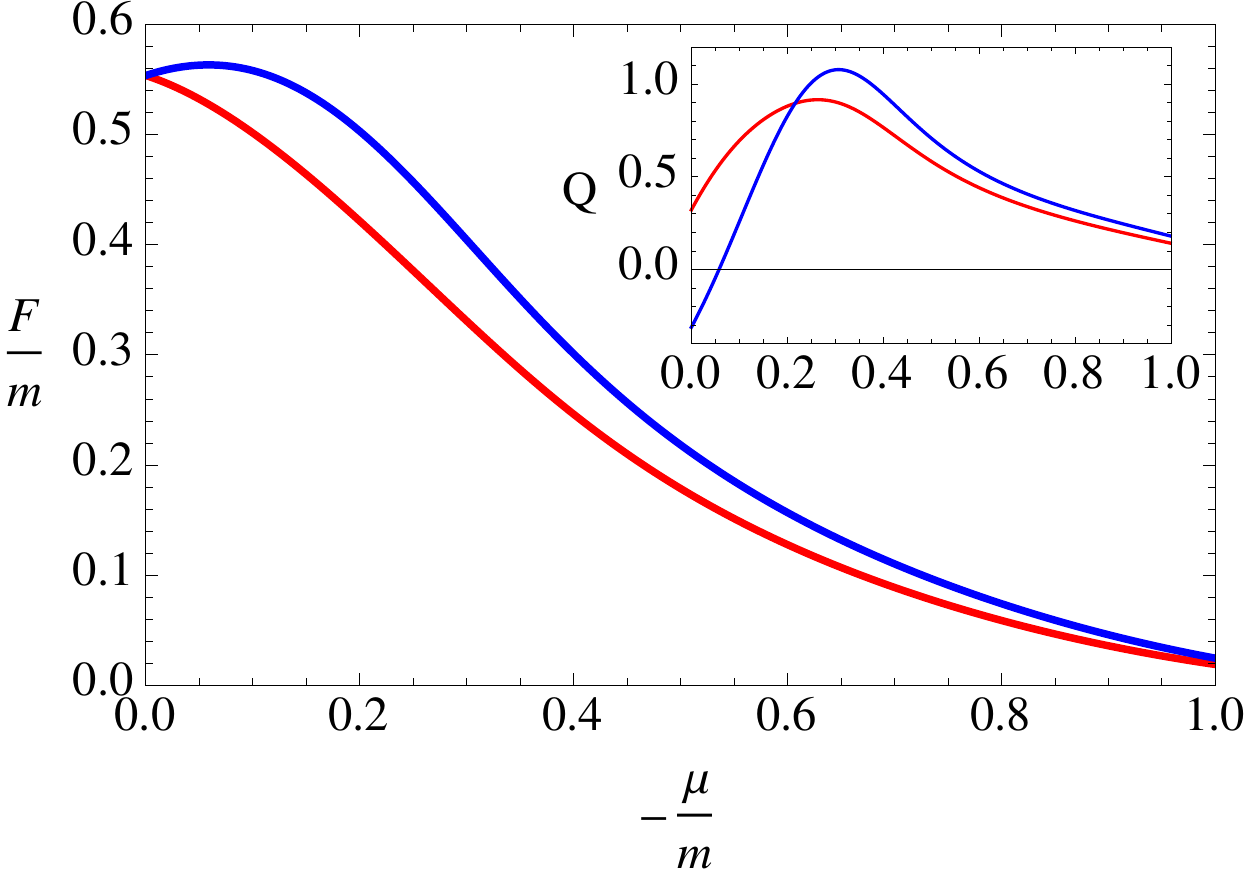}
	\caption{\label{fig:PressureEntropy2} The Polyakov loops $\ell$ (blue) and $\bar\ell$ (red) and the corresponding free energies as functions of $\hat\mu=-\mu$. The insets in the first plot shows a close-up view on the small $\hat\mu$ region where the change of monotony for $\ell$ (and therefore $F_q$) occurs. The second inset shows the average baryonic charge of the thermal and chemical bath in the presence of a test quark (blue) or a test anti-quark (red).}
\end{figure}

Finally, in Fig.~\ref{fig:PressureEntropy2}, we show the Polyakov loops and the corresponding free energies as functions of\footnote{Our choice of for the sign of the term $\propto\mu$ in \Eqn{eq:action} in unconventional and can be interpreted by saying that the usual chemical potential associated to the baryonic charge is $\hat\mu=-\mu$.} $\hat\mu=-\mu$ for fixed $T/m=0.33$ and $M/m=2.22$, for $N_f=3$ degenerate flavors (just at one-loop, since the two-loop result suffers from the limitation that we described above). We observe that the Polyakov loops have a different monotony at small $\hat\mu$ but then increase together towards one, just as was observed in Ref.~\cite{Dumitru:2003hp}. In this reference, the different monotony at small $\hat\mu$ was used to question the interpretation of the logarithms of the Polyakov loops as free energies. We now show that this statement is not necessarily true if the charge of the bath at $\hat\mu=0$ is not zero. Moreover, using a simple thermodynamical argument, we show that the interpretation of the logarithms of the Polyakov loops as free energies leads precisely to the qualitative behavior observed for their $\mu$-dependence.\footnote{This does not prove that the logarithms of the Polyakov loops can be interpreted in terms of free energies, but merely that this interpretation cannot be ruled out using the non-monotonous behavior with respect to $\mu$.}

To this purpose, let us first consider the free energy of the bath $F=-T\ln {\rm tr}\,\exp\{-\beta(H-\hat\mu Q)\}$, where $Q$ is the baryonic charge and $\hat\mu=-\mu$ reflects our unconventional choice for the sign of the term $\propto\mu$ in \Eqn{eq:action}. One easily obtains that
\beq
\frac{\partial F}{\partial\hat\mu} = -\langle Q \rangle \quad {\rm and} \quad \frac{\partial\langle Q\rangle}{\partial\hat\mu}=\beta \left<(Q-\langle Q\rangle)^2\right>>0\,.
\eeq
Now, in absence of any external sources, the thermal bath is charge-conjugation invariant for $\hat\mu=0$, from which it follows that $\langle Q\rangle_{\hat\mu=0}=0$. The above inequalities then imply that $\langle Q\rangle>0$ and $\frac{\partial F}{\partial\hat\mu}<0$  for $\hat\mu>0$: the free energy of the bath is a decreasing function of $\hat\mu$. This conclusion, however, gets modified in presence of external sources. Consider, in particular, the bath in presence of a static quark ($q$) or antiquark ($\bar q$): because the test charge breaks the charge-conjugation invariance, one has  $\langle Q\rangle_{q,\hat\mu=0}=-\langle Q\rangle_{\bar q,\hat\mu=0}\neq0$. The test particles being static, they only interact in the color singlet channel with the particles of the bath. This interaction being attractive, the presence of a static quark (resp. anti-quark) has the tendency to bring a negative (resp. positive) baryonic charge to the bath: $\langle Q\rangle_{q,\hat\mu=0}<0$ and $\langle Q\rangle_{\bar q,\hat\mu=0}>0$. The equations above then imply that
\beq
\forall \hat\mu>0\,,\quad \langle Q\rangle_{\bar q}>0\,,
\eeq
while there exists a certain $\hat\mu_0>0$ such that,\footnote{$\langle Q\rangle_{\bar q}$ and $\langle Q\rangle_{\bar q}$ should be of the same sign at very large $\mu$.}
\beq
\forall \hat\mu\in [0,\hat\mu_0]\,,\,\,\langle Q\rangle_{q}<0 \quad {\rm and} \quad \forall \hat\mu>\hat\mu_0\,,\,\,\langle Q\rangle_{q}>0\,.
\eeq
In terms of the free energies, this implies that $F_{\bar q}$ is monotonously decreasing for $\hat\mu>0$, while $F_{q}$ first increases and then decreases. This is  shown in Fig.~\ref{fig:PressureEntropy2} and is exactly what was observed in Ref.~\cite{Dumitru:2003hp}.

\subsection{Thermodynamical stability}

Thermodynamical stability can be studied by analyzing the shape of the function $s(e,q)$ where $s$ is the entropy density, $e$ the energy density and $q$ the charge density. Let us recall how this comes about using the pure YM limit as an example, in which case we can drop the dependence with respect to $q$ and study the curve $s(e)$.

Suppose that the system is put in some box of volume $\Omega$ and let us denote by $S=\Omega\,s$ and $U=\Omega\,e$ the total entropy and total internal energy respectively. The total entropy is an extensive function of both the energy and the volume $S(U,\Omega)=\Omega\,s(U/\Omega)$. The system will be stable if any redistribution of the internal energy in the volume $\Omega$ leads to a decrease of the total entropy. This means that
\beq
S(U_1,x_1\Omega)+S(U_2,x_2\Omega)\leq S(U_1+U_2,\Omega)\,,
\eeq
for any $ U_1,U_2$ such that $U_1+U_2=U$ and $0<x_1,x_2<1$, with  $x_1+x_2=1$.  This is equivalent to
\beq
S(x_1U_1,x_1\Omega)+S(x_2U_2,x_2\Omega)\leq S(x_1U_1+x_2U_2,\Omega)\,,
\eeq
which, using the extensivity of $S(U,\Omega)$ and introducing $e_i=U_i/\Omega$, rewrites as
\beq\label{eq:stabil}
x_1s(e_1)+x_2s(e_2)\leq s(x_1e_1+x_2e_2)\,,
\eeq
with $x_1e_1+x_2e_2=e$.

\begin{figure}[t]
	\centering
	\includegraphics[width=0.42\textwidth]{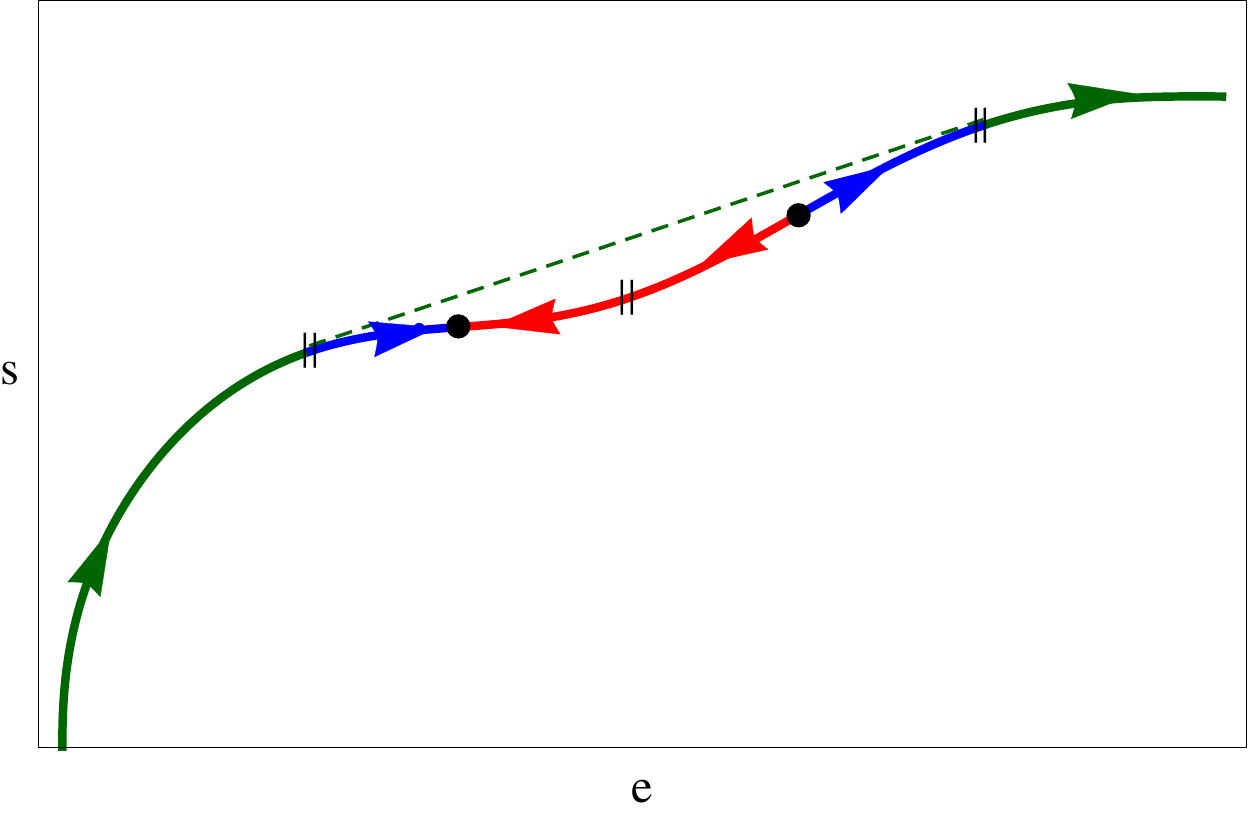}
	\caption{Typical shape of the function $s(e)$, in the presence of a first order phase transition. The green curve, including the dashed line, corresponds to the convex envelope $\bar s(e)$. It departs from $s(e)$ at two points represented by a $\parallel$ with slope corresponding to (the inverse of) the transition temperature. There is a third point in the convex region of the curve (in red) which has the same temperature (inverse slope). The concave and convex regions are separated by the spinodal points represented by a dot in the figure. The arrows indicate how the system evolves as one increases the temperature (decreases the slope). At low enough temperature, there is only one possible point on the curve (left green branch). As the temperature increases, there is a point on the left green portion whose temperature equals the rightmost spinodal. At this temperature a metastable and unstable state appear on the rightmost spinodal in addition to the stable state. When the system reaches the transition temperature, it can either jump directly to the (stable) right green branch or continue in the form of a metastable state along the (metastable) left blue branch until this state coalesces with the unstable one.
\label{fig:firstorder}}
\end{figure}

To interpret this identity, let us assume first that $s(e)$ is a concave function. In this case, it is clear that the stability condition (\ref{eq:stabil}) is fulfilled for any value of $e$. Indeed, for any interval $[e_1,e_2]$ containing $e$, the entropy curve over this interval lies always above the line that connects $(e_1,s(e_1))$ to $(e_2,s(e_2))$.  When the entropy curve is not concave, it is convenient to introduce the convex envelope $\bar s(e)$. The typical shape is shown in Fig.~\ref{fig:firstorder} (the curve should be monotonously increasing since $ds/de=1/T$). At a given $e$, such that $s(e)=\bar s(e)$, we have stability since the condition (\ref{eq:stabil}) is always satisfied. If $s(e)\neq\bar s(e)$ but $s$ is still concave around $e$, we have a metastable state, since small redistributions of the energy are still compatible with Eq.~(\ref{eq:stabil}). Finally, if $s$ is convex in the vicinity of $e$, we have an unstable state, since any redistribution of the energy violates the above inequality and allows to increase the entropy. The above is typical of a first order phase transition, the transition occurring at the points where $s(e)$ departs from $\bar s(e)$, and the inflexion points of $s(e)$ corresponding to the spinodal points. In terms of the extrema of the effective potential, the branches where $s(e)=\bar s(e)$ correspond to the the absolute minimum, the concave branches such that $s(e)\neq\bar s(e)$ correspond to the local minima and the convex branch to the maximum.

\begin{figure}[t]
	\centering
	\includegraphics[width=0.44\textwidth]{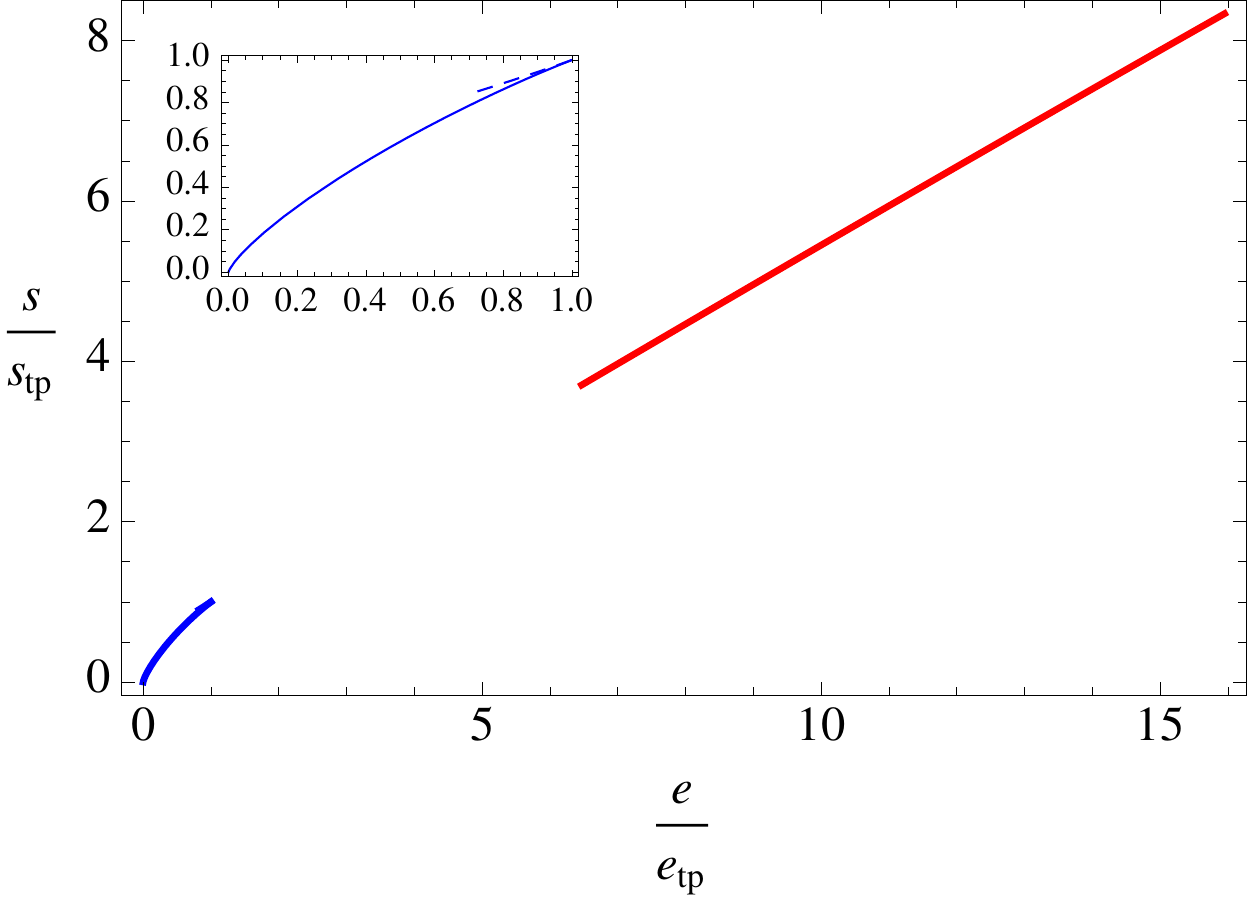}
	\vglue5mm
		\includegraphics[width=0.44\textwidth]{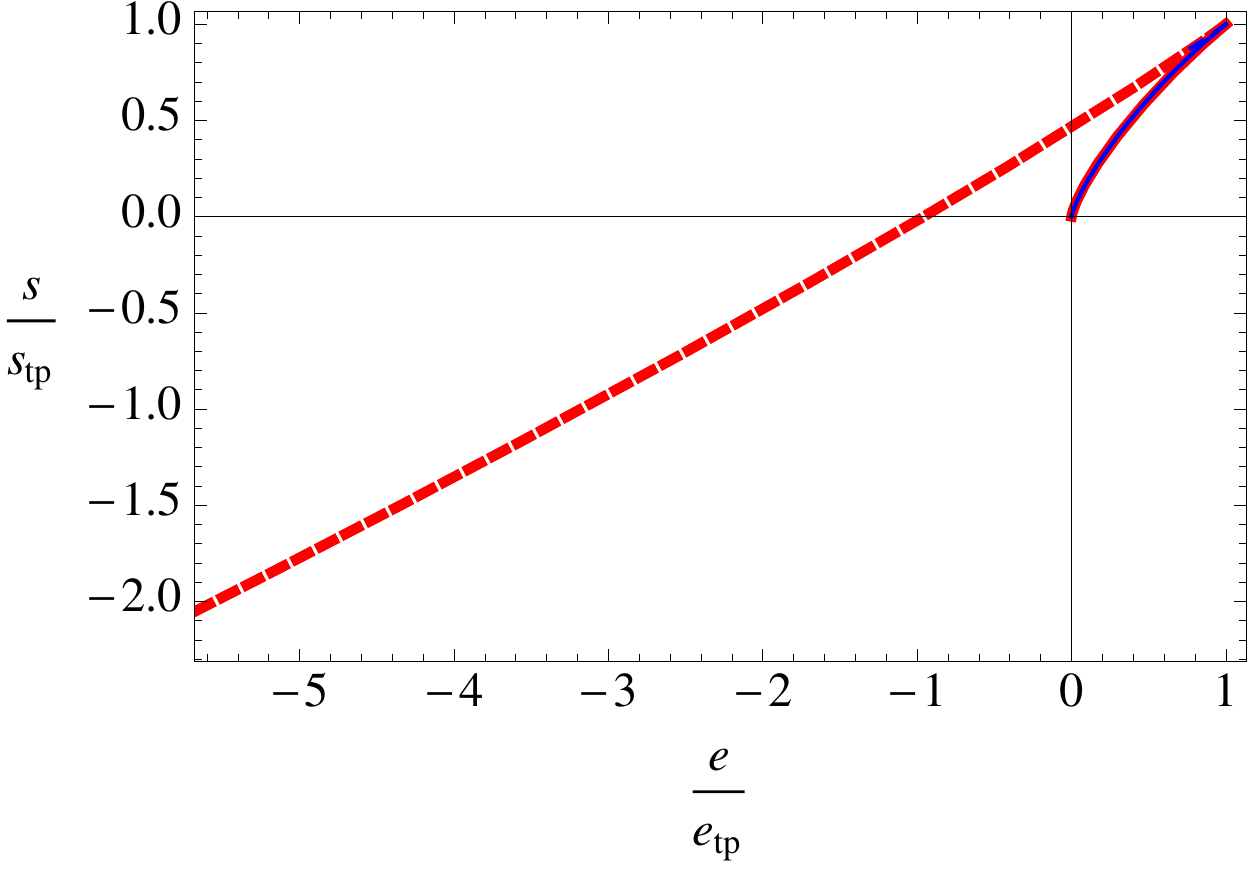}
	\caption{Top: entropy density vs. energy density at two-loop at $\mu=0$. Both quantities are normalized by their respective values $s_{\rm tp}$ and $e_{\rm tp}$ at the turning point. The discontinuity corresponds  to the first order phase transition. The non-concave branch of $s(e)$ is represented by a dashed line. The insert is a zoom on the low temperature (blue) branch. Bottom: comparison of the low temperature branch at one- (red) and two-loop (blue) orders. The unphysical part (dashed lines) is significantly reduced at two-loop order. In particular, neither $s$ nor $e$ explore negative values. \label{fig:stability}}
\end{figure}

We can evaluate the relation $s(e)$ in the present model at one- and two-loop orders. The corresponding curves for $s(e)$ are shown in Fig.~\ref{fig:stability}. We show only the branch corresponding to the absolute minimum. We note first that, even though the entropy density can become negative in the one-loop case when approaching the transition temperature, it is positive at all temperatures in the two-loop case. However, both at one- and two-loop orders, we observe that the function $s(e)$ is not univalued and, after the turning point, the curve $s(e)$ becomes convex, thus indicating an instability. It is important to mention that the turning point does not correspond to a spinodal and that the convex branch is obtained from the absolute minimum of the potential. Therefore this instability is not the one inherent to the first order phase transition. Instead, this unstable convex multivalued branch is characterized by\footnote{$de/dT$ and $ds/dT$ have always the same sign since $de/dT=Tds/dT$. In the case where they are both $>0$, this implies that $s(e)$ is a univalued function. For an infinite system, we can have $dT/de=0$ in some range of $e$, when a first order transition is present. However this does not spoil the argument that $s(e)$ should be univalued.} $de/dT<0$ and $ds/dT<0$. That $de/dT$ is negative contradicts the general formula
\beq\label{eq:U}
U=\frac{1}{Z}{\rm tr}\,H\,e^{-\beta H}\,,
\eeq
which leads to
\beq
\frac{\partial U}{\partial T}=\beta^2\langle(H-U)^2\rangle\,.
\eeq
This could signal that, in the phenomenological model that we are considering for the Gribov completion of the Landau gauge-fixing, the partition function may still involve contributions from negative norm states \cite{deBoer:1995dh}. We mention however that the non-concave region appears in the vicinity of $T_c$ and not $\smash{T=0}$, and that it shrinks from { $\Delta T_{\rm 1\ell}/m=0.133$ at one-loop order to $\Delta T_{\rm 2\ell}/m=0.095$} at two-loop order, which could also mean that this unstable branch is a spurious effect of the perturbative expansion. As a final remark, we note that $dp/de=dp/dT\times dT/de=sdT/de$. The entropy density being positive (at two-loop order), the change of sign in $de/dT$ also turns $dp/de<0$ and violates Le Chatelier principle. Similar violations have been observed within the Gribov-Zwanziger approach \cite{Canfora:2016xnc}.

A similar discussion holds in the presence of a conserved charge but now in terms of the function $s(e,\rho)$ where $\rho$ is the charge density.

	\section{Conclusion}\label{sec:conc}
	
We have extended a previous perturbative investigation \cite{reinosa2015perturbative} of the phase diagram of QCD with heavy quarks in the context of a massive extension of background field methods by including two-loop corrections to the background field effective potential. In particular, we have computed here the two-loop quark sunset diagram, to be added to the pure glue two-loop contributions obtained in Ref.~\cite{reinosa2016two}, for a general class of gauge groups. We have explicitly studied the phase structure of the $SU(3)$ theory with various (heavy) quark contents at nonzero temperature and (real or imaginary) chemical potential. Our main conclusion is that, when properly interpreted, the two-loop results quantitatively improve the one-loop ones for what concerns the phase diagram. For instance the ratio of critical masses to critical temperatures in the Columbia plot, known from lattice simulations, are very well reproduced. These results add to the accumulating list of (seemingly nonperturbative) infrared properties of non-Abelian gauge theories that are efficiently captured by a modified perturbation theory in terms of a simple massive Lagrangian in this class of gauges \cite{Tissier:2010ts,Tissier:2011ey,Pelaez:2013cpa,Reinosa:2013twa,Reinosa:2014ooa,Reinosa:2014zta,reinosa2016two,reinosa2015perturbative,Reinosa:2017qtf,Pelaez:2017bhh}.

We have also analyzed the effects of two-loop corrections to thermodynamical observables, which suffer from spurious artifacts at one-loop order. As was the case for the pure YM theories \cite{Reinosa:2014zta,reinosa2016two}, we find that most of these spurious one-loop features---such as, e.g., the negative entropy and pressure or the thermodynamic instability---are either completely or strongly washed out at two-loop order. The main open question in this context remains the spurious contribution of massless modes to thermodynamics at low temperature. These play an essential role in getting the correct phase diagram, in particular, the existence of a confined phase. However they should not contribute to actual thermodynamic observables. This is a central issue, not only of the present perturbative approach, but of all existing continuum methods. 

Finally, we have shown, using simple thermodynamical arguments, that the behavior of the Polyakov loops as functions of the chemical potential agrees with their interpretation in terms of quark and anti-quark free energies, as one expects from general arguments \cite{Svetitsky:1985ye}.\\
	
\acknowledgements{We thank M.~Pel\'aez, J.~M. Pawlowski, M.~Tissier and N.~Wschebor for useful discussions related to this work.}\\

\appendix

\section{Reduction to scalar-type integrals}\label{Reduction}
Here, we outline how to obtain the expression in Eq.~(\ref{ScalarInts}) in terms of the scalar tadpoles and the scalar sunset introduced in Eqs.~(\ref{bosonicTads}), (\ref{fermionicTads}), and (\ref{scalarSunset})  from the original expression for the quark sunset in Eq.~(\ref{Quarksunset}). The first step is to deal with the Dirac structure, for which we note the general formulae
\begin{eqnarray}
\gamma_\mu (i  \slashed{P}+M )\gamma_\mu &=& (2-d) i  \slashed{P}+ d\, M\,,\nonumber\\
\slashed{Q}(i  \slashed{P}+M) \slashed{Q}&=& Q^2(-i \slashed{P}+M) + 2i (P\cdot Q) \slashed{Q}\,,\nonumber 
\end{eqnarray}
which lead to the following expression for the quark sunset contribution to the background field effective potential:
\begin{widetext}
\begin{eqnarray}
&& V_q^{(2)}(r,T,\mu)=\frac{g^2}{2} {\rm tr} \mathds{1}  \sum_f \sum_{\sigma \rho \kappa} {\cal D}_{\sigma,\rho \kappa} \Bigg \{ - (d-1)\int_{\hat P} \int_Q  \Big[P^\rho \cdot L^\sigma +M_f^2  \Big]  G_{m}(Q^\kappa)G_{M_f}(P^\rho) G_{M_f}(L^\sigma) \nonumber \\ 
&& \hspace{3.5cm}+\,2\int_{\hat P} \int_Q  \Big[ Q^2_\kappa P^\rho \cdot L^\sigma - (P^\rho \cdot Q^\kappa)(Q^\kappa \cdot L^\sigma)\Big]   G_{0}(Q^\kappa) G_{m}(Q^\kappa)G_{M_f}(P^\rho) G_{M_f}(L^\sigma) \Bigg \}.
\end{eqnarray}
Here, we have combined some of the terms that arise from the various products of gamma matrices. The combination under the square brackets in the second line can be written in various ways using the transverse projector $P^\perp_{\mu\nu}(Q^\kappa)\equiv\delta_{\mu\nu}-Q^\kappa_\mu Q^\kappa_\nu/Q_\kappa^2$ and $L^\sigma=P^\rho+Q^\kappa$. For instance,
\beq
Q^2_\kappa (P^\rho \cdot L^\sigma)-(P^\rho \cdot Q^\kappa)(Q^\kappa \cdot L^\sigma) = Q_\kappa^2\,P^\rho\cdot P^\perp(Q^\kappa)\cdot L^\sigma= Q_\kappa^2\,P^\rho\cdot P^\perp(Q^\kappa)\cdot P^\rho=P_\rho^2Q_\kappa^2-(P_\rho\cdot Q_\kappa)^2 \,.
\eeq
Upon substitution, this results in
\begin{eqnarray}\label{appeq:sisi}
&&V_q^{(2)}(r,T,\mu)=\frac{g^2}{2} {\rm tr} \mathds{1}  \sum_f \sum_{\sigma \rho \kappa} {\cal D}_{\sigma,\rho \kappa} \Bigg \{ - (d-1) \int_{\hat P} \int_Q  \Big[  P^\rho \cdot L^\sigma +M_f^2  \Big]  G_{m}(Q^\kappa)G_{M_f}(P^\rho) G_{M_f}(L^\sigma) \nonumber \\ 
& & \hspace{3.5cm}+\,2 \int_{\hat P} \int_Q  \Big[ P^2_\rho Q^2_\kappa - (P^\rho \cdot Q^\kappa)^2 \Big]   G_{0}(Q^\kappa) G_{m}(Q^\kappa)G_{M_f}(P^\rho) G_{M_f}(L^\sigma) \Bigg \}.
\end{eqnarray}
We can now rewrite the bracket of the first line as
\beq
P^\rho \cdot L^\sigma +M_f^2 = \frac{1}{2}\big[P_\rho^2+L_\sigma^2-Q_\kappa^2\big]+M_f^2=  \frac{1}{2} \big[L^2_\sigma +M_f^2 + P^2_\rho  + M_f^2 -Q^2_\kappa -m^2 \big] + \frac{m^2}{2}\,.
\eeq
To treat the integrals in the second line of \Eqn{appeq:sisi} we further use $G_{0}(Q^\kappa)G_{m}(Q^\kappa) = [ G_{0}(Q^\kappa) - G_{m}(Q^\kappa)]/m^2$. Altogether, we get
\begin{equation} \label{ReduceIntermediate}
V_q^{(2)}(r,T,\mu)=- \frac{g^2}{4} {\rm tr} \mathds{1}  \sum_f \sum_{\sigma \rho \kappa} {\cal D}_{\sigma,\rho \kappa} \Bigg \{  (d-1) \Big[ \big( J^\rho_{M_f} + J^\sigma_{M_f} \big) J^\kappa_m - J^\rho_{M_f}J^\sigma_{M_f} +m^2 S_{mM_fM_f}^{\kappa\rho\sigma}\Big] + \frac{4}{m^2}\Big[ I^{\kappa\rho\sigma}_{mM_fM_f}- I^{\kappa\rho\sigma}_{0M_fM_f} \Big]  \Bigg\},
\end{equation}
where we have used the definitions of the tadpoles and scalar sunset given in Eqs.~(\ref{bosonicTads}), (\ref{fermionicTads}) and (\ref{scalarSunset}), as well as
\begin{eqnarray}
I^{\kappa\rho\sigma}_{mMM}\equiv\int_{\hat P} \int_Q \Big[ P^2_\rho Q^2_\kappa - (P^\rho \cdot Q^\kappa)^2 \Big]  \, G_{m}(Q^\kappa)\, G_{M}(P^\rho) \, G_{M}(L^\sigma)\,. 
\end{eqnarray}
The final step in the reduction to scalar integrals is to simplify $ I^{\kappa\rho\sigma}_{mMM}$. To this end we rewrite 
\begin{eqnarray}
P^2_\rho Q^2_\kappa G_{m}(Q^\kappa)\, G_{M}(P^\rho) \, G_{M}(L^\sigma) &=& G_{M}(L^\sigma)-\left[ m^2 G_{m}(Q^\kappa)+M^2G_{M}(P^\rho)\right] G_{M}(L^\sigma) \nonumber \\
&+&\,m^2 M^2 \, G_{m}(Q^\kappa) G_{M}(P^\rho) G_{M}(L^\sigma)\,,
\end{eqnarray}
and, similarly,
\begin{eqnarray}
&&(P^\rho \cdot Q^\kappa)^2  G_{m}(Q^\kappa)\, G_{M}(P^\rho) \, G_{M}(L^\sigma)\nonumber \\ 
&& \hspace{0.5cm}=\,\frac{m^2}{2}  (P^\rho \cdot Q^\kappa) G_{m}(Q^\kappa) G_{M}(P^\rho) G_{M}(L^\sigma)\nonumber \\  
&& \hspace{1.0cm}+\,\frac{(P^\rho \cdot Q^\kappa) }{2} \big[ G_{m}(Q^\kappa)  G_{M}(P^\rho) - G_{m}(Q^\kappa) G_{M}(L^\sigma) -   G_{M}(P^\rho)G_{M}(L^\sigma)   \big]      \nonumber \\
&& \hspace{0.5cm}=\,\frac{m^4}{4} G_{m}(Q^\kappa) G_{M}(P^\rho) G_{M}(L^\sigma)  + \frac{m^4}{4} \big[G_{m}(Q^\kappa) G_{M}(P^\rho) -  G_{m}(Q^\kappa) G_{M}(L^\sigma) - G_{M}(P^\rho) G_{M}(L^\sigma)\big]  \nonumber \\ 
&& \hspace{1.0cm}+\,\frac{(P^\rho \cdot Q^\kappa)}{2}\big[ G_{m}(Q^\kappa) G_{M}(P^\rho) - G_{m}(Q^\kappa)G_{M}(L^\sigma) - G_{M}(P^\rho)G_{M}(L^\sigma)\big].
 \end{eqnarray}  
This then leads to
\begin{eqnarray}
  I^{\kappa\rho\sigma}_{mMM}=-\frac{1}{2} \big[\tilde{J}_m^\kappa (\tilde{J}_{M}^\rho -\tilde{J}_{M}^\sigma) - \tilde{J}_{M}^\rho\tilde{J}_{M}^\sigma \big] - \frac{m^2}{4} J_m^\kappa (J^\rho_{M} + J^\sigma_{M}) + \frac{m^2-2M^2}{4} J^\rho_{M}J^{\sigma}_{M}+ m^2 \left(M^2 - \frac{m^2}{4}\right) S_{mMM}^{\kappa\rho\sigma},
\end{eqnarray} 
where $S_{mMM}^{\kappa\rho\sigma}$ has been defined in \Eqn{scalarSunset}. Using this expression for $ I^{\kappa\rho\sigma}_{mMM}$ in Eq.~(\ref{ReduceIntermediate}) reproduces Eq.~(\ref{ScalarInts}).\\
\end{widetext}

\section{Splitting in terms of thermal factors}
\label{ThermalFactors}
It is convenient to decompose the scalar integrals defined above in contributions with a given number of thermal (Bose-Einstein or Fermi-Dirac) factors. For the bosonic tadpoles, we refer for instance to Ref.~\cite{reinosa2016two} and simply quote here the results: $J_m^\kappa=J_m^\kappa(0n)+J_m^\kappa(1n)$ with
\begin{eqnarray}
J_m^\kappa(0n) & = & \int_q\frac{1}{2\varepsilon_q}\equiv J_m(0n)\,,\\
J_m^\kappa(1n) & = &  \int_q \frac{n_{\varepsilon_{m,q}-i\hat{r}\cdot\kappa}+n_{\varepsilon_{m,q}+i\hat{r}\cdot\kappa}}{2\varepsilon_{m,q}}\,,
\end{eqnarray}
where $\smash{n_x\equiv(e^{\beta x}-1)^{-1}}$ is the Bose-Einstein distribution function, $\varepsilon_{m,q}\equiv\sqrt{q^2+m^2}$ and $\int_q\equiv\mu_r^{2\epsilon}\int\frac{d^{d-1}q}{(2\pi)^{d-1}}$, with $\mu_r$ the renormalization scale. We have also introduced $\hat r\equiv Tr$. 
Similarly, $\tilde J_m^\kappa=\tilde J_m^\kappa(0n)+\tilde J_m^\kappa(1n)$, with 
\beq
 \tilde J_m^\kappa(0n)=0
\eeq
 and
\beq	\label{appeq:tildeJmb}
\tilde{J}_m^\kappa(1n)=\int_q \frac{n_{\varepsilon_{m,q}-i\hat{r}\cdot\kappa}-n_{\varepsilon_{m,q}+i\hat{r}\cdot\kappa}}{2i}\,.
\eeq
We mention however that, in contrast to Ref.~\cite{reinosa2016two}, we have not written $J_m^\kappa(1n)$ and $\tilde J_m^\kappa(1n)$ as real and imaginary parts respectively. This is because, as already explained in Sec.~\ref{sec:add_quarks}, for real chemical potentials, the background component $r_8$ needs to be continued from $\mathds{R}$ to $i\mathds{R}$. Irrespectively of these considerations, we have $J_m^{-\kappa}=J_m^\kappa$ and $\tilde J_m^{-\kappa}=-\tilde J_m^\kappa$.
	
We recall finally that $J_m(0n)$ can be computed in $d=4-2\epsilon$ dimensions as
\beq	\label{appeq:Jm}
J_m(0n)=-\frac{m^2}{16 \pi^2}\bigg[ \frac{1}{\epsilon} + \rm{ln}\, \frac{\bar{\mu}^2}{m^2} +1 + {\cal O}(\epsilon) \bigg],
\eeq
where $\bar{\mu}^2\equiv 4 \pi e^{-\gamma}\mu^2$, with $\gamma$ the Euler constant.

\subsection{Fermionic Tadpoles $J^\rho_M$ and $\tilde{J}^\rho_M$}
	
Similar decompositions hold for the fermionic tadpoles.
Standard contour integration techniques lead to $J_{M}^\rho=J_{M}^\rho(0n)+J_{M}^\rho(1n)$, with
\begin{eqnarray}
J^\rho_M(0n)&=&  J_M(0n)\,,\\
\label{appeq:Jm1n}J^\rho_M(1n)&=&  -\int_p \frac{f_{\varepsilon_{M,p}-i\rho \hat{r} + \mu}+\,f_{\varepsilon_{M,p}+i\rho \hat{r} - \mu}}{2 \varepsilon_{M,p}} \,,\nonumber\\
\end{eqnarray}
where $f_x\equiv(e^{\beta x}+1 )^{-1}$ is the Fermi-Dirac distribution. 
	
The second fermionic tadpole can be treated in analogy with the bosonic case, see Ref.~\cite{Reinosa:2014zta}, where it was noted that, although the corresponding Matsubara sum is not absolutely convergent, it  be defined as the limit of the symmetric sum: $\lim_{N\to\infty}\sum_{-N}^{N}$. We extend the sum by a term that vanishes due to the symmetry under $P_0\to-P_0$ as 
\beq
\sum_{-N}^{N}  P^\rho_0 G_{M}(P^\rho) = \sum_{-N}^{N} \big[  P^\rho_0 G_{M}(P^\rho) -P_0 G_{M}(P)|_{\mu=0}  \big]  
\eeq
and then apply standard contour integration to find
\beq\label{eq:u1}
\tilde{J}^\rho_{M} = -\int_p \frac{f_{\varepsilon_{M,p}-i\rho \hat{r} + \mu} - f_{\varepsilon_{M,p}+i\rho \hat{r} - \mu}}{2i}\,.  
\eeq
As in the bosonic case, there is no zero thermal factor contribution for this tadpole integral, $\tilde{J}^\rho_{M}(0n)=0$, so that $\tilde{J}^\rho_{M}=\tilde{J}^\rho_{M}(1n)$. Note that $J_{M}^{\rho}(\mu)=J_{M}^{-\rho}(-\mu)$ and $\tilde{J}^{\rho}_{M}(\mu) = -\tilde{J}^{-\rho}_{M}(-\mu)$.\\
     
	\subsection{Scalar Sunset $S_{mMM}^{\kappa\rho\sigma}$}
	
	For a similar calculation, see Ref.~\cite{Marko:2010cd}. We start by rewriting the scalar sunset 
	\beq
		S_{mMM}^{\kappa\rho\sigma}=\int_{\hat P}\int_Q   G_m(Q^\kappa)G_{M}(P^\rho)G_{M}(L^\sigma)
	\eeq
	by making use of the spectral representations
	\begin{eqnarray}
G_m(Q^\kappa) & = & \int\frac{dq_0}{2\pi}\frac{\rho_m(q_0,q)}{q_0-i\omega^\kappa_n}\,,\\
G_{M}(P^\rho) & = & \int\frac{dp_0}{2\pi}\frac{\rho_{M}(p_0,p)}{p_0-i\hat\omega^\rho_n}\nonumber\\
 & = & \int\frac{dp_0}{2\pi}\frac{\rho_{M}(p_0,p)}{p_0+i\hat\omega^\rho_n}\,,
\end{eqnarray}
with $\rho_\alpha(q_0,q)= 2\pi \, \varepsilon(q_0)\,  \delta(q_0^2-\varepsilon^2_{\alpha,q})$
	and the shifted bosonic and fermionic Matsubara frequencies
	\begin{eqnarray}
	\omega_n^\kappa &=& \omega_n +\hat{r}\cdot\kappa\,,\\
	\hat{\omega}_n^\rho &=& \hat{\omega}_n +\hat{r}\cdot\rho +i\mu\,.
	\end{eqnarray}
	We obtain
	\begin{widetext}
	\begin{eqnarray}
		S_{mMM}^{\kappa\rho\sigma} &=& T^2 \int_{p}\int_q \sum_n \sum_m \int_{q_0}\frac{\rho_{m}(q_0,q)}{q_0-i\omega_n^\kappa} \int_{p_0}\frac{\rho_{M}(p_0,p)}{p_0-i\hat{\omega}_m^\rho}
        \int_{l_0}\frac{\rho_{M}(l_0,l)}{l_0+i\hat{\omega}_{n+m}^\sigma}. 
	\end{eqnarray}
	The full ($\sigma$) and partial ($\sigma_m$) Matsubara sums, defined as 
	\begin{eqnarray}
	\sigma \equiv T \sum_m \sigma_m = T^2 \sum_{n,m} \frac{1}{(q_0-i\omega_n^\kappa)(p_0-i\hat{\omega}_m^\rho)(l_0+i\hat{\omega}_{n+m}^\sigma)}.
	\end{eqnarray} are easily computed. We find
	\begin{eqnarray}
	\sigma_m = \frac{1}{p_0-i\hat{\omega}_m^\rho}\frac{1}{q_0+l_0+ i\hat{\omega}_m^\rho}\big[ n_{q_0-i\hat{r}\cdot\kappa} + f_{-l_0-i \hat{r}\cdot\sigma  +\mu}  \big], \nonumber
	\end{eqnarray}
where we have used $n_{\varepsilon+i\hat\omega_m}=-f_\varepsilon $, and  
\begin{eqnarray}
\sigma = \frac{1}{q_0+p_0+l_0}\big[ - f_{p_0-i\hat{r}\cdot\rho +\mu} + f_{ -l_0-q_0-i\hat{r}\cdot\rho +\mu}  \big] \, \big[ n_{q_0-i\hat{r}\cdot\kappa} + f_{-l_0 -i \hat{r}\cdot\sigma +\mu} \big]. \nonumber
\end{eqnarray}	
Note that both the denominator and the numerator vanish  in the limit $q_0+p_0+l_0\rightarrow 0$ and $\sigma$ remains well defined. To simplify the following steps, it is useful to rewrite the above expression such that each thermal factor depends on only one of $\{l_0,p_0,q_0 \}$. This is achieved by using the identity $ f_{x+y}\big( 1+n_x -f_y \big)=n_xf_y$. We have 
\begin{eqnarray}
\sigma =  -\frac{1}{q_0+p_0+l_0}\bigg[ n_{-q_0+i\hat{r}\cdot\kappa}  f_{-l_0-i \hat{r}\cdot\sigma +\mu} + f_{p_0-i \hat{r}\cdot\rho + \mu}  \big( n_{q_0-i\hat{r}\cdot\kappa} + f_{-l_0-i \hat{r}\cdot\sigma  +\mu} \big) \bigg].
\end{eqnarray} 
In order to extract the divergent contributions, it is convenient to write, for $q_0$ and $p_0\neq 0$,
\begin{eqnarray}
n_{q_0-i\hat{r}\cdot\kappa} &=& -\theta(-q_0)+\varepsilon(q_0) \,n_{|q_0|-\varepsilon(q_0) i\hat{r}\cdot\kappa}\,,\\
 f_{p_0-i \hat{r}\cdot\rho + \mu}&=& \theta(-p_0)+ \varepsilon(p_0) \,f_{ |p_0| +\varepsilon(p_0) (\mu-i \hat{r}\cdot\rho)}\,,
\end{eqnarray}
where $\theta(x)$ and $\varepsilon(x)$ denote the Heaviside and sign functions respectively.\footnote{ We stress that $f_{ |p_0| +\varepsilon(p_0) (\mu-i \hat{r}\cdot\rho)}$ still contains zero-temperature contributions, which, however, lead to ultraviolet-finite integrals. The same applies to $n(|q_0|-\varepsilon(q_0) i\kappa \hat{r})$ when certain components of the background are taken imaginary.} When splitting the thermal factors in this fashion, the a priori well-defined expression for $\sigma$ above picks up divergences at $ p_0+l_0+q_0=0$ since now not all numerators equally vanish. We therefore regularize the denominator as was done in Ref.~\cite{Reinosa:2014zta} by 
\beq
\frac{1}{p_0+l_0+q_0}\rightarrow {\rm Re}\, \frac{1}{p_0+l_0+q_0 +i0^{+}}\,.
\eeq 

We obtain
$S_{mMM}^{\kappa\rho\sigma} =\, S_{mMM}^{\kappa\rho\sigma}(0n)+S_{mMM}^{\kappa\rho\sigma}(1n)+S_{mMM}^{\kappa\rho\sigma}(2n)$. The contribution with no thermal factors depends neither on the background nor on the temperature or the chemical potential. It can thus be disregarded for our purpose here.
The contribution with one thermal factor is found to be
\begin{eqnarray}
S_{mMM}^{\kappa\rho\sigma}(1n)= \left[ J_{M}^{\rho}(1n) + J_{M}^{\sigma}(1n)\right] \, {\rm Re}\,  I_{Mm}(\varepsilon_{M,p}+i0^{+},p) + J_{m}^{\kappa}(1n) \, {\rm Re}\,  I_{MM}(\varepsilon_{m,q}+i0^{+},q)\,,
\end{eqnarray}
where we have introduced $I_{\alpha \beta}(z;k)=I_{\beta \alpha}(z;k)$ as 
\begin{eqnarray}
I_{\alpha \beta}(z;k)&=& \int_{q,q_0,l_0} \rho_{\alpha}(q_0,q) \rho_\beta(l_0,q+k) \, \frac{\theta(l_0) -\theta(-q_0)}{q_0+l_0+z},
\end{eqnarray}
which is the analytic continuation of the following $T=0$ integral:
\beq
I_{\alpha \beta}(K)\equiv I_{\alpha\beta}(i\omega;k)=\mu^{2\epsilon}\int\frac{d^dQ}{(2\pi)^d} G_\alpha(Q)G_\beta(Q+K)\,,
\eeq
 in the (complex) frequency plane; see Refs.~\cite{Blaizot:2004bg,Reinosa:2014zta} for more details. This facilitates the evaluation of the quantities $I_{Mm}(0n) \equiv {\rm Re}\,  I_{Mm}(\varepsilon_{M,p}+i0^{+},p)$ and $I_{MM}(0n) \equiv {\rm Re}\,  I_{MM}(\varepsilon_{m,q}+i0^{+},q)$. One finds, with $x\equiv m^2/M^2$,
\begin{eqnarray}\label{appeq:IMm}
I_{Mm}(0n) \, 
&=& \frac{1}{16\pi^2}\left[\frac{1}{\epsilon} + \ln \, \frac{\bar{\mu}^2}{M^2}+2  - \frac{x^2}{2}\, \ln \, x^2 
+ x^2\sqrt{\left|1-\frac{4}{x^2}\right|}\begin{cases}
- \, { \mathrm{arctan} } \, \sqrt{\frac{4}{x^2} -1 }    & 0 <x^2 < 4  \\
\,  \quad \mathrm{artanh} \, \sqrt{1-\frac{4}{x^2}}  & x^2 \geq 4
\end{cases}\right]\,,
\end{eqnarray} 
and
\begin{eqnarray} \label{appeq:IMM2}
I_{MM}(0n) \, 
&=& 
\frac{1}{16\pi^2}\left[\frac{1}{\epsilon} + \ln \, \frac{\bar{\mu}^2}{M_{f}^2} + 2 
- 2  \, \, \sqrt{ \left| 1- \frac{4}{x^2}\right|  }  \quad
\begin{cases} 
\, \mathrm{arctan} \,\, \sqrt{\frac{1}{\frac{4}{x^2} -1}  }  & 0<x^2 <4 \\
\, \mathrm{artanh} \,\,  \sqrt{1- \frac{4}{x^2}}     & x^2 \geq 4
\end{cases}\right].
\end{eqnarray}
Finally, collecting all terms with two thermal factors we find 
\begin{eqnarray}
S_{mMM}^{\kappa\rho\sigma}(2n)&=& - \int_{p_0,p} \rho_{M}(p_0,p)\varepsilon(p_0)  f_{|p_0| +\varepsilon(p_0) (\mu-i \hat{r}\cdot\rho)}  \int_{q_0,q} \rho_{m}(q_0,q)\varepsilon(q_0) n_{|q_0| -\varepsilon(q_0) i\hat{r}\cdot\kappa}   \nonumber \\
		&  & \hspace{1.4cm}\left.\times\, {\rm Re}\, G_{M}(p_0+q_0+i0^{+}, |p+q|)    \right. \nonumber\\
&-&   \int_{p_0,p} \rho_{M}(p_0,p)\varepsilon(p_0)  f_{ |p_0| +\varepsilon(p_0) (\mu-i \hat{r}\cdot\sigma)}  \int_{q_0,q} \rho_{m}(q_0,q)\varepsilon(q_0) n_{|q_0| -\varepsilon(q_0) i\hat{r}\cdot\kappa}   \nonumber \\
&  & \hspace{1.4cm}\left.\times\, {\rm Re}\, G_{M}(p_0+q_0+i0^{+}, |p+q|)    \right. \nonumber\\
&+&   \int_{p_0,p} \rho_{M}(p_0,p)\varepsilon(p_0)  f_{|p_0| +\varepsilon(p_0) (\mu-i \hat{r}\cdot\rho)}  \int_{q_0,q} \rho_{M}(q_0,q)\varepsilon(q_0)  f_{|q_0| +\varepsilon(q_0) (\mu-i \hat{r}\cdot\rho)}    \nonumber \\
&  & \hspace{1.4cm}\left.\times\, {\rm Re}\, G_{m}(p_0+q_0+i0^{+}, |p+q|)    \right. .
\end{eqnarray}
Integrating over the frequencies and the angles we end up with
\begin{eqnarray}\label{appeq:S2n}
S_{mMM}^{\kappa\rho\sigma}(2n)&=&- \frac{1}{64\pi^4}\sum_{\lambda,\tau\in\{+,-\}} \int_0^\infty dp \frac{p}{\varepsilon_{M,p}} \big( f_{\varepsilon_{M,p} +\lambda (\mu-i \hat{r}\cdot\rho)}  + f_{\varepsilon_{M,p} -\lambda (\mu-i \hat{r}\cdot\sigma)} \big) \int_0^\infty dq \frac{q}{ \varepsilon_{m,q}} n_{\varepsilon_{m,q}-i\tau \hat{r}\cdot\kappa} \nonumber \\
&  & \hspace{3.0cm}\left.\times\, {\rm Re}\, \ln \, \frac{\varepsilon_{M,p+q}^2-\big(\lambda \varepsilon_{M,p} +\tau \varepsilon_{m,q} +i0^{+} \big)^2 }{\varepsilon_{M,p-q}^2-\big(\lambda \varepsilon_{M,p} +\tau \varepsilon_{m,q} +i0^{+} \big)^2}    \right.  \nonumber \\
&+& 
\frac{1}{64\pi^4}\sum_{\lambda,\tau\in\{+,-\}} \int_0^\infty dp \frac{p}{\varepsilon_{M,p}}  f_{\varepsilon_{M,p} +\lambda (\mu-i \hat{r}\cdot\rho)}  \int_0^\infty dq\frac{q}{\varepsilon_{M,q}}  f_{\varepsilon_{M,q} -\tau (\mu-i \hat{r}\cdot\sigma)} \nonumber \\
&  & \hspace{3.0cm}\left.\times\, {\rm Re}\, \ln \, \frac{\varepsilon_{m,p+q}^2-\big(\lambda \varepsilon_{M,p} +\tau \varepsilon_{M,q} +i0^{+} \big)^2 }{\varepsilon_{m,p-q}^2-\big(\lambda \varepsilon_{M,p} +\tau \varepsilon_{M,q} +i0^{+} \big)^2}    \right.\,.
\end{eqnarray}
Expressions \eqn{appeq:tildeJmb}, \eqn{appeq:Jm}, \eqn{appeq:Jm1n}, \eqn{eq:u1}, \eqn{appeq:IMm}, \eqn{appeq:IMM2}, and \eqn{appeq:S2n} are to be used in \Eqn{eq:res} to compute the complete potential at two-loop order.

\end{widetext}

\section{Large $ \mu-T$ expansion}\label{sec:last}
		
We now consider the behavior of $V_q^{(2)}(r,T,\mu)$ as $\mu,T \rightarrow  \infty$ while keeping $\hat{\mu}=\frac{\mu}{T}$ fixed. The leading contribution will be  $\sim T^4$. We shall thus consider the limit of $V_q^{(2)}(r,T,\mu)/T^4$ as a function of $\hat\mu$. 

We start determining the behavior or each of the scalar integrals. The bosonic tadpoles $J_m^\kappa(1n)$ and $\tilde{J}^\kappa_m(1n)$  do not depend on $\mu $, so their leading behaviour coincides with that obtained as $T\rightarrow \infty$. The calculation follows the steps detailed in Ref.~\cite{Reinosa:2014zta} and we find
\begin{eqnarray}
	J^\kappa_m(1n)\sim\frac{T^2}{2\pi^2} P_2(r\cdot\kappa)\,, \quad \tilde{J}^\kappa_m \sim \frac{T^3}{2\pi^2} P_3(r\cdot\kappa)\,,
\end{eqnarray} 
where the $P_n$'s denote the integrals
\begin{eqnarray}
\label{appeq:Podd}
P_{2n+1}(z) & \equiv & \int_0^\infty dx\,x^{2n}\,\frac{\tilde n_{x-iz}-\tilde n_{x+iz}}{2i}\,,\\
\label{appeq:Peven}
P_{2n+2}(z) & \equiv & \int_0^\infty dx\,x^{2n+1}\,\frac{\tilde n_{x-iz}+\tilde n_{x+iz}}{2}\,,
\end{eqnarray}
with $\tilde n$ the Bose-Einstein distribution function with $\beta$ set to $1$. Contrary to Ref.~\cite{Reinosa:2014zta}, we have not written these integrals as imaginary and real parts respectively, since we shall also consider them in the case where $z$ is complex. In this way, the above expressions are analytic everywhere, except when $z$ is a multiple of $2\pi$. In particular, the functions $P_n(z)$ are analytic over the simply connected open sets $z\,\in\,]2\pi n,2\pi(n+1)[\times \mathds{R}$, seen as subsets of $\mathds{C}$. For real $x$ in the interval $]0,2\pi[$\,, they are given by polynomials \cite{Dumitru:2013xna,Reinosa:2014zta}:
\begin{align}
\label{appeq:P1}
P_1(x) & = \frac{\pi-x}{2}\,,\\
P_2(x) & = \frac{(\pi-x)^2}{4}-\frac{\pi^2}{12}\,,\\
P_3(x) & = -\frac{(\pi-x)^3}{6}+\frac{\pi^2 (\pi-x)}{6}\,,\\
P_4(x) & = -\frac{(\pi-x)^4}{8}+\frac{\pi^2 (\pi-x)^2}{4}-\frac{7\pi^4}{120}\,,\,\dots\label{appeq:P4}
\end{align}It follows that the functions $P_n(z)$ are given by the same polynomial expressions on the whole stripe $]0,2\pi[\times \mathds{R}$.\footnote{Another possibility is to compute $P_n(r)$ explicitly for any $r\in\mathds{C}$. First, the $P_n$'s are related to each other by some recursion relations, obtained in Ref.~\cite{Reinosa:2014zta} for real $r$ but which are in fact valid for $r\in\mathds{C}$. These relations allow to generate all the $P_n$'s by successive integrations of $P_1(r)$. Expanding the thermal factor, one then finds that $P_1(r)=\sum_{k=-\infty}^\infty (1-\delta_{k0})e^{ikr}/2ik=(1/2i)\ln[(1+e^{-i(r-\pi)})/(1+e^{i(r-\pi)})]=(1/2i)\ln e^{-i(r-\pi)}$ which equals $(\pi-r)/2$ if the real part of $r$ lies between $0$ and $2\pi$.} To obtain them in any other stripe $]2\pi n,2\pi(n+1)[\,\times\,\mathds{R}$, one uses the periodicity property $P_n(z+2\pi)=P_n(z)$. {We mention finally that, except for $P_1$, the functions $P_n$ are continous which means that the above polynomial expressions can be extended to the boundary of the stripe $]0,2\pi[\times \mathds{R}$, in particular for $r=0$ or $r=2\pi$. This is not the case for $P_1(x)$ which is discontinous at these points. However, as we will check below, this has no influence on the expression for the total potential and for convenience we shall also define $P_1(0)$ and $P_1(2\pi)$ from the polynomial expression.}

For later purpose it is convenient to introduce similar integrals for fermions:
\begin{eqnarray}
\label{appeq:hatPodd}
\hat P_{2n+1}(z) & \equiv & -\int_0^\infty dx\,x^{2n}\,\frac{\tilde f_{x+iz}-\tilde f_{x-iz}}{2i}\,,\\
\label{appeq:hatPeven}
\hat P_{2n+2}(z) & \equiv & -\int_0^\infty dx\,x^{2n+1}\,\frac{\tilde f_{x+iz}+\tilde f_{x-iz}}{2}\,.
\end{eqnarray}
Owing to the identity $\tilde f_x=-\tilde n_{x\pm i\pi}$, we have the relation
\beq\label{eq:pi}
\hat P_n(z)=P_n(z+\pi)\,.
\eeq
In particular, the $\hat P_n$'s are polynomials in the stripe $]-\pi,\pi[\,\times\,\mathds{R}$. Another formula follows from the identity $\tilde f_x=\tilde n_x-2\tilde n_{2x}$, namely,
\beq
\hat P_n(z)=\frac{1}{2^{n-1}}P_n(2z)-P_n(z)\,.
\eeq
We immediately deduce that, in the limit $T\to\infty$ with $\hat\mu$ fixed,
\beq
J^\rho_M(1n)\sim\frac{T^2}{2\pi^2} \hat P_2(r\cdot\rho+i\hat\mu)\,,  \quad \tilde{J}^\rho_M \sim \frac{T^3}{2\pi^2} \hat P_3(r\cdot\rho+i\hat\mu)\,.
\eeq
By rescaling the integration momenta by $T$ it is easily argued that $S_{mMM}^{\kappa\rho\sigma}(2n)/T^4\to 0$ as $T\to\infty$.\\

We now have all the ingredients to find a closed expression for $\lim_{T\to\infty} V_q^{(2)}(r,T,\mu)/T^4$. The only contributions that survive in this limit are products of tadpoles with one thermal factor. From $[J^\rho_{M_f}(1n)+J^\sigma_{M_f}(1n)]J^\kappa_m(1n)$, we obtain
\beq
\frac{1}{4\pi^4}{P_2}(r\cdot\kappa)\Big[\hat P_2(r\cdot\rho+i\hat\mu) +\hat P_2(r\cdot\sigma+i\hat\mu)\Big],
\eeq 
and, from $-J^\rho_{M_f}(1n)J^\sigma_{M_f}(1n)$,
\beq
-\, \frac{1}{4 \pi^4}\hat P_2(r\cdot\rho+i\hat\mu) \hat P_2(r\cdot\sigma+i\hat\mu)\,.           
\eeq
The products involving tadpoles of the second type seem dominant. However they appear in the combination $[\tilde J_0^\kappa(1n)-\tilde J_m^\kappa(1n)][\tilde J_M^\rho(1n)-\tilde J_M^\sigma(1n)]/m^2$. The dominant $T^3$ behavior of $\tilde J_0$ and $\tilde J_m$ cancels in the difference and is replaced by $Tm^2$, which together with the factor $1/m^2$ and the factor $\tilde J_M^\rho$ yields terms of the same order as products of tadpoles of the first type. More precisely, as $T\to\infty$ and fixed $\hat\mu$,
\beq
\frac{\tilde J_0^\kappa-\tilde J_m^\kappa}{m^2} \sim  -\left.\frac{d\tilde J_m^\kappa}{dm^2}\right|_{m^2=0}=\frac{T}{4\pi^2}P_1(r\cdot\kappa)\,,
\eeq
where we have used the identities 
\beq
\int_Q \omega_n G(Q^\kappa)^2=-\int_Q \frac{\omega_n}{2q}\frac{d}{dq}G(Q^\kappa)=\int_Q \frac{\omega_n}{2q^2}G(Q^\kappa)
\eeq
 as well as the same Matsubara sum as  in \Eqn{eq:u1}. The term $[\tilde J_0^\kappa(1n)-\tilde J_m^\kappa(1n)][\tilde J_M^\rho(1n)-\tilde J_M^\sigma(1n)]/m^2$ then behaves as
\beq
\frac{{P_1}(r\cdot\kappa)}{8\pi^4} \, \Big[\hat P_3(r\cdot\rho+i\hat\mu)-\hat P_3(r\cdot\sigma+i\hat\mu)      \Big].
\eeq
Combining all these pieces as in Eq.~(\ref{eq:res}), we arrive at (a summation over fermionic flavours is implied)
\begin{eqnarray} 
&& \lim_{T\to\infty}\frac{V_q^{(2)}(r,T,\mu)}{T^4}=-\frac{g^2}{2\pi^4}\sum_{\sigma \rho \kappa} {\cal D}_{\sigma , \rho \kappa }\nonumber\\
&& \hspace{1.0cm} \times\,\bigg[-\hat P_2(r\cdot\rho+i\hat\mu) \hat P_2(r\cdot\sigma+i\hat\mu)\nonumber\\
&& \hspace{1.5cm}+\,2{P_2}(r\cdot\kappa) \hat P_2(r\cdot\rho+i\hat\mu)\nonumber\\
&& \hspace{1.5cm}+\,{P_1}(r\cdot\kappa)\hat P_3(r\cdot\rho+i\hat\mu) \big)    \bigg],
\end{eqnarray}
where we have used ${\cal D}_{\sigma,\rho\kappa}={\cal D}_{\rho,\sigma(-\kappa)}$ and the fact that the functions $P_{2n}$ and $P_{2n+1}$ are even and odd respectively. 

As for the one-loop quark contribution, we have, from Ref.~\cite{reinosa2015perturbative}, $V_q^{(1)}(r,T,\mu)=\sum_{f,\rho}V_f^0(T,\mu-ir_\rho T)$, where
\beq
V_{f}^0(T,\mu) = -\frac{1}{3\pi^2}\!\int_0^\infty \!\!dp\,\frac{p^4}{\varepsilon_{M_f,p}}\left[f_{\varepsilon_{M_f,p}+\mu}+(\mu\to-\mu)\right].
\eeq
At large $T$ and fixed $\hat\mu$, the fermion masses are irrelevant and we find
\beq
\lim_{T\to\infty}\frac{V_q^{(1)}(r,T,\mu)}{T^4}=\frac{2N_f}{3\pi^2}\sum_\rho \hat P_4(r\cdot\rho+i\hat\mu)\,.
\eeq

Finally, the pure glue contributions have been obtained before \cite{Weiss:1980rj,Gross:1980br,reinosa2016two} as
\beq
\lim_{T\to\infty}\frac{V_g^{(1)}(r,T)}{T^4}=-\frac{1}{3\pi^2}\sum_\kappa P_4(r\cdot\kappa)
\eeq
for the one-loop part and
\begin{eqnarray}
& & \lim_{T\to\infty}\frac{V_g^{(2)}(r,T)}{T^4}=\frac{g^2}{4\pi^4}\sum_{\kappa\lambda\tau}{\cal C}_{\kappa\lambda\tau}\Big[P_2(r\cdot \kappa)P_2(r\cdot \lambda)\nonumber\\
& & \hspace{3.5cm}+\,P_1(r\cdot\kappa)P_3(r\cdot\lambda)\Big]
\end{eqnarray}
for the two-loop term.
Note the analogies between the pure glue and the quark contribution formulae. We also mention that these expressions can be written in more symmetric forms by making use of the properties of the tensors ${\cal C}_{\kappa\lambda\tau}$ and ${\cal D}_{\sigma,\rho\kappa}$ as well as of the parity properties of the functions $P_n$ and $\hat P_n$. It is convenient to separate the cases where $\kappa$, or related indices, can be a zero or a root, just as in Ref.~\cite{reinosa2016two}. After some algebra, we find

\begin{widetext}
\begin{eqnarray}
\label{appeq:poilocu}
{ \lim_{T\to\infty}\frac{V_g(r,T)}{T^4}} & = & {\frac{1}{3\pi^2}\left[d_C P_4(0)+2{\sum_\alpha}^* P_4(r\cdot\alpha)\right]}\nonumber\\
& + & \frac{g^2}{2\pi^4}{\sum_\alpha}^* \alpha^2\Big\{\big[2P_2(0)+P_2(r\cdot \alpha)\big]P_2(r\cdot \alpha)-P_1(r\cdot\alpha)P_3(r\cdot\alpha)\Big\}\nonumber\\
& + & \frac{g^2}{\pi^4}{\sum_{\alpha\beta\gamma}}^*{\cal C}_{\alpha\beta\gamma}\Big[P_2(r\cdot \alpha)P_2(r\cdot \beta)+P_2(r\cdot \beta)P_2(r\cdot \gamma)+P_2(r\cdot \gamma)P_2(r\cdot \alpha)\Big]\nonumber\\
& + & \frac{g^2}{2\pi^4}{\sum_{\alpha\beta\gamma}}^*{\cal C}_{\alpha\beta\gamma}\Big[P_1(r\cdot\alpha)P_3(r\cdot\beta)+P_1(r\cdot\beta)P_3(r\cdot\gamma)+P_1(r\cdot\gamma)P_3(r\cdot\alpha)\nonumber\\
& & \hspace{2.0cm}+P_3(r\cdot\alpha)P_1(r\cdot\beta)+P_3(r\cdot\beta)P_1(r\cdot\gamma)+P_3(r\cdot\gamma)P_1(r\cdot\alpha)\Big],
\end{eqnarray}
with $d_C$ the dimension of the Cartan subalgebra and where $\sum^*_\alpha$ is a sum over the pairs of roots $(\alpha,-\alpha)$ and $\sum^*_{\alpha\beta\gamma}$ sums over the pairs of triplets $((\alpha,\beta,\gamma),(-\alpha,-\beta,-\gamma))$ such that $\alpha+\beta+\gamma=0$, with the possible permutations of $(\alpha, \beta, \gamma)$ being counted only once. For the quark contribution, we find, similarly,
\begin{eqnarray} 
\label{appeq:poilone}
{\lim_{T\to\infty}\frac{ V_q(r,T,\mu)}{T^4}} & = & {\frac{2N_f}{3\pi^2}\sum_\rho \hat P_4(r\cdot\rho+i\hat\mu)}\nonumber\\
& - & \frac{g^2}{2\pi^4}\sum_\rho \rho^2\Big[2{P_2}(0)-\hat P_2(r\cdot\rho+i\hat\mu)\Big] \hat P_2(r\cdot\rho+i\hat\mu)\nonumber\\
& - & \frac{g^2}{\pi^4}{\sum_{\sigma \rho \alpha}}^* {\cal D}_{\sigma , \rho \alpha }\Big[{P_2}(r\cdot\alpha)\hat P_2(r\cdot\rho+i\hat\mu)-\hat P_2(r\cdot\rho+i\hat\mu) \hat P_2(r\cdot\sigma+i\hat\mu)+{P_2}(r\cdot\alpha)\hat P_2(r\cdot\sigma+i\hat\mu)\Big]\nonumber\\
& - & \frac{g^2}{2\pi^4}{\sum_{\sigma \rho \alpha}}^* {\cal D}_{\sigma , \rho \alpha }\Big[{P_1}(r\cdot\alpha)\hat P_3(r\cdot\rho+i\hat\mu)-{P_1}(r\cdot\alpha)\hat P_3(r\cdot\sigma+i\hat\mu)  \Big]\,,
\end{eqnarray}
\end{widetext}
where $\sum^*_{\sigma \rho \alpha}$ sums the pairs of triplets $((\sigma,\rho,\alpha),(\rho,\sigma,-\alpha))$ such that $\sigma=\rho+\alpha$.

Finally, we stress that the polynomial expressions Eqs.~\eqn{appeq:P1}--\eqn{appeq:P4} of the functions $P_n$ and, using \Eqn{eq:pi}, of the functions $\hat P_n$ { require the scalar products $r\cdot\alpha$ and $r\cdot\rho$ to lie respectively in the intervals $\in [0,2\pi]$ and $[-\pi,\pi]$.} Using these expressions thus requires one { to subtract enough multiples of $2\pi$ to $r\cdot\alpha$ and $r\cdot\rho$, which leaves the integrals \eqn{appeq:Podd}, \eqn{appeq:Peven}, \eqn{appeq:hatPodd}, and \eqn{appeq:hatPeven} invariant. In fact, it is not so difficult to know a priori how many multiples need to be subtracted, as we now explain for the SU($3$) case.} As far as the $r\cdot\alpha$'s are concerned, using the parity properties of the polynomials, we can always rewrite the above expressions such that only the scalar products $r\cdot\alpha^{(j=1,2,3)}$ occur, with $\alpha^{(1)}=(1,\sqrt{3})/2$, $\alpha^{(2)}=(1,-\sqrt{3})/2$ and $\alpha^{(3)}=(1,0)$. Decomposing the background as $r=4\pi x_j \rho^{(j)}$, where $\rho^{(1)}=(1,1/\sqrt{3})/2$ and $\rho^{(2)}=(1,-1/\sqrt{3})/2$ are weights of ${\bf 3}$ and ${\bf \bar 3}$, respectively, the conditions $0<r\cdot\alpha^{(1,2)}<2\pi$ become $0<x_{1,2}<1$ since $\alpha^{(j)}\cdot \rho^{(k)}=\delta_{jk}/2$ for $j,k\in\{1,2\}$. Because $4\pi\rho^{(1)}$ and $4\pi\rho^{(2)}$ are two of the edges of the {\it fundamental} Weyl chamber $\{(0,0),2\pi(1,1/\sqrt{3}),2\pi(1,-1/\sqrt{3})\}$, these conditions are satisfied over a region twice as big as this Weyl chamber. The third condition $0<r\cdot\alpha^{(3)}<2\pi$ is only fulfilled over half of this region, that is the fundamental Weyl chamber itself. { We conclude that the use of the polynomials \eqn{appeq:P1}--\eqn{appeq:P4} is valid over the whole Weyl chamber for the pure YM contribution \eqn{appeq:poilocu} provided that we rewrite it in terms of the scalar products $r\cdot\alpha^{(1)}$, $r\cdot\alpha^{(2)}$ and $r\cdot\alpha^{(3)}$.}

\begin{center}
\begin{figure}[b]  
\epsfig{file=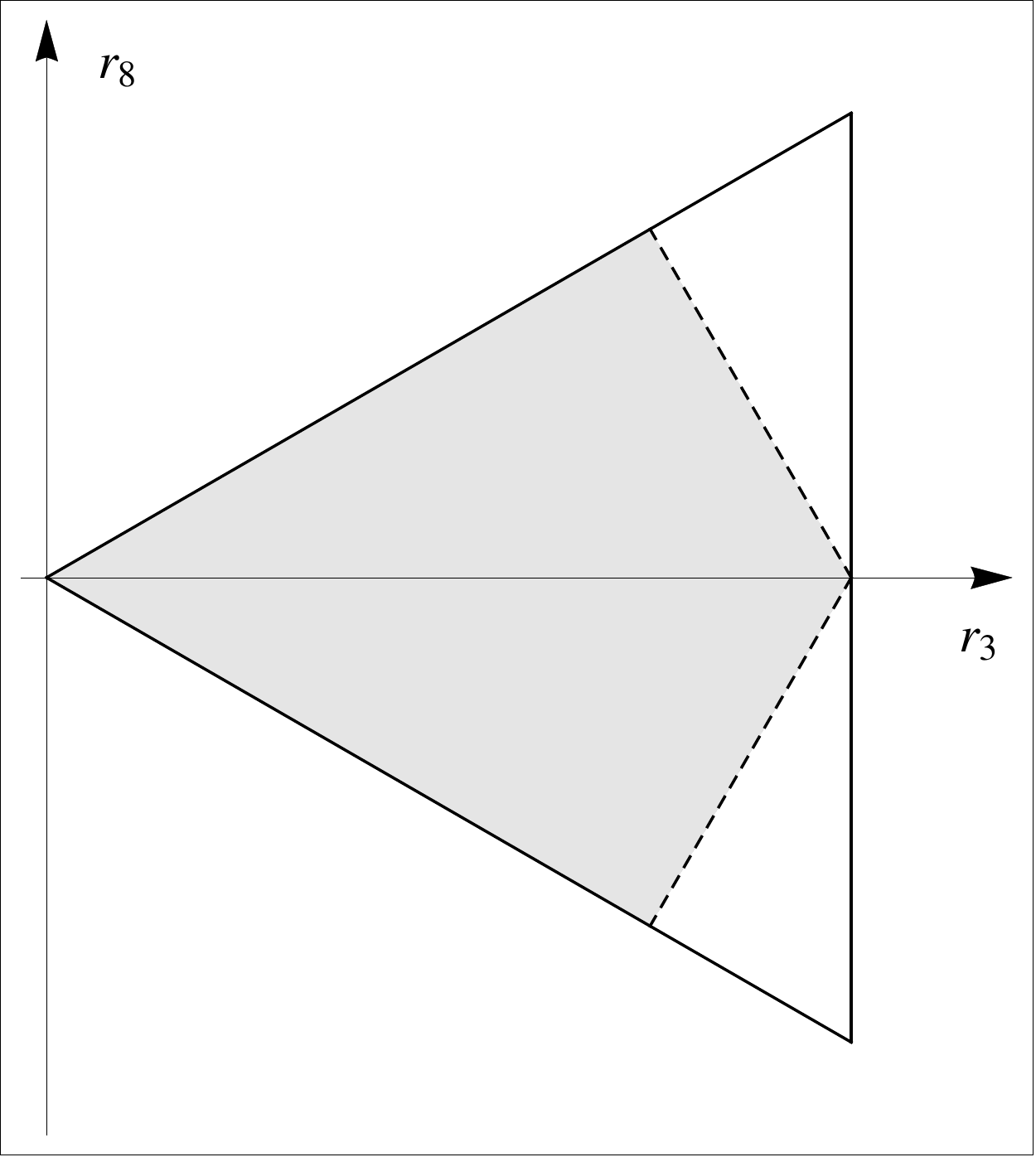,width=5.7cm}
\caption{The shaded region represents the region of the fundamental Weyl chamber (equilateral triangle) where one can use the polynomials \eqn{appeq:P1}--\eqn{appeq:P4} and \eqn{eq:pi}, provided one has rewritten the formula for the potential in such a way that only occur scalar products of the form $r\cdot\alpha^{(j)}$ and $r\cdot\rho^{(j)}$, with the $\alpha^{(j)}$'s and $\rho^{(j)}$'s defined in the main text. In the pure YM case, one can use the polynomials (\ref{appeq:P4}) over the whole fundamental Weyl chamber.}\label{fig:region}
\end{figure}
\end{center}

A similar discussion can be made for the products $r\cdot\rho$. First it is always possible to rewrite our expressions in terms of $r\cdot\rho^{(j=1,2,3)}$, where $\rho^{(1,2)}$, have been defined above and $\rho^{(3)}=(0,1/\sqrt{3})$. The conditions $-\pi<r\cdot\rho^{(1,2)}<\pi$ translate into $-1/2<y_{1,2}<1/2$, where the $y_j$ are the coordinates of $r$ in the basis $\{4\pi\alpha^{(1)},4\pi\alpha^{(2)}\}$. In the fundamental Weyl chamber, this delimitates the shaded region shown in Fig.~\ref{fig:region}. Away from this region one has to shift either $r\cdot\rho^{(1)}$ or $r\cdot\rho^{(2)}$ by $-2\pi$ before using the polynomial. Finally, the constraint $-\pi<r\cdot \rho^{(3)}<\pi$ is satisfied over the whole fundamental Weyl chamber.

{ As mentioned above, the potential never involves the value of $P_1(r)$ at $r=2\pi n$. Suppose that $r$ is such that $r\cdot\alpha=2\pi n$, then for $\alpha+\beta+\gamma=0$, we have $r\cdot\beta=-2\pi n-r\cdot\gamma$ and therefore the contributions $P_1(r\cdot\alpha)P_3(r\cdot\alpha)$ and $P_1(r\cdot\alpha)(P_3(r\cdot\beta)+P_3(r\cdot\gamma))$ to Eq.~(\ref{appeq:poilocu}) vanish (because $P_3$ is periodic and odd), irrespectively of the value of $P_1(2\pi n)$. Similarly, for $\sigma=\rho+\alpha$, we have $r\cdot\sigma=r\cdot\rho+2\pi n$ and therefore the contribution ${P_1}(r\cdot\alpha)\hat P_3(r\cdot\rho+i\hat\mu)-{P_1}(r\cdot\alpha)\hat P_3(r\cdot\sigma+i\hat\mu)$ to Eq.~(\ref{appeq:poilone}) vanishes, irrespectively of the value of $P_1(2\pi n)$.}

In the case of a vanishing chemical potential, the relevant part of the Weyl chamber is the axis $r_8=0$. In that case, we obtain, up to corrections ${\cal O}(g^4)$
\begin{widetext}
\beq
\label{appeq:Vglue}
\left. \lim_{T\to \infty}\frac{V_g(r,T)}{T^4}\right|_{r_8=0}=\frac{135 r_3^4-600 \pi r_3^3+720 \pi ^2 r_3^2-256 \pi ^4}{1440\pi ^2}
+g^2\frac{189 r_3^4-912 \pi r_3^3+1584 \pi ^2 r_3^2-1152 \pi ^3 r_3+256 \pi ^4}{1536\pi ^4},
\eeq
and
\beq
\label{appeq:Vquark}
 \left.\lim_{T\to \infty}\frac{V_q(r,T,\mu=0)}{T^4}\right|_{r_8=0}=-\frac{5r_3^4- 40 \pi^2 r_3^2+56 \pi^4 }{480\pi ^2} -g^2\frac{111 r_3^4-264 \pi r_3^3-96 \pi ^2 r_3^2+576\pi ^3 r_3-320 \pi ^4}{4608\pi ^4}\,.
\eeq
The total potential for $N_f=1$ reads
\beq
\label{appeq:Vtotal}
\left.\lim_{T\to \infty}\frac{V(r,T,\mu=0)}{T^4}\right|_{r_8=0}=\frac{15r_3^4-75\pi  r_3^3+105\pi^2r_3^2-53\pi^4}{180\pi ^2}
+g^2\frac{57 r_3^4-309 \pi r_3^3+606 \pi ^2 r_3^2-504 \pi ^3 r_3+136 \pi ^4}{576\pi ^4}\,.
\eeq
\end{widetext}
The minimum of the potential \eqn{appeq:Vtotal} is given by
\beq
 r_3^{\rm min}=\frac{3g^2}{4\pi}+{\cal O}(g^4).
\eeq
Interestingly, it is the same as the minimum of the pure Yang-Mills potential \eqn{appeq:Vglue}, which implies that the quarks do not modify the high-temperature value of the Polyakov loop at ${\cal O}(g^2)$ for $\mu=0$. We mention that the formulas (93) and (94) of Ref.~\cite{reinosa2016two} contain an error which has been corrected here.

\end{document}